\documentclass[useAMS,usenatbib]{mn2e}

\usepackage[T1]{fontenc}
 \usepackage{pslatex}
\usepackage{amsmath}
\usepackage{amsfonts}
\usepackage{xcolor}
\usepackage{url}
\usepackage{bm}
\usepackage{graphicx}
\usepackage{amssymb}
\usepackage{soul}
\usepackage{booktabs}
\usepackage{array}
\usepackage{ulem}
\usepackage[utf8]{inputenc}
\inputencoding{latin1}
\inputencoding{utf8}

\newcommand{\quotes}[1]{``#1''}

\renewcommand{\vec}[1]{\boldsymbol{#1}}
\def\lsim{ \lower .75ex\hbox{$\sim$} \llap{\raise .27ex \hbox{$<$}} }
\def\gsim{ \lower .75ex \hbox{$\sim$} \llap{\raise .27ex \hbox{$>$}} }

\def\lsim{ \lower .75ex\hbox{$\sim$} \llap{\raise .27ex \hbox{$<$}} }
\def\gsim{ \lower .75ex \hbox{$\sim$} \llap{\raise .27ex \hbox{$>$}} }
\title[] 
{Evolution of current  and pressure driven instabilities \\ in relativistic jets}

\author[M. Musso et al.] {M. Musso$^{1}$, G. Bodo$^{2}$\thanks{E-mail:
gianluigi.bodo@inaf.it}, G. Mamatsashvili$^{3,4}$, P. Rossi$^{2}$ and A. Mignone$^{1}$\\
$^{1}$Dipartimento di Fisica,  Universit\`a degli Studi di Torino, Via Pietro Giuria 1, 10125 Torino, Italy\\
$^2$INAF -- Osservatorio Astrofisico di Torino, Strada Osservatorio 20, I-10025 Pino Torinese, Italy\\
$^{3}$Helmholtz-Zentrum Dresden-Rossendorf, Bautzner Landstraße 400, D-01328 Dresden, Germany\\
$^{4}$Abastumani Astrophysical Observatory, Abastumani 0301, Georgia
}

\begin{document}

\maketitle

\begin{abstract} 
Instabilities in relativistic magnetized jets are thought to be deeply connected to their energy dissipation properties and to the consequent acceleration of the non-thermal emitting relativistic  particles. Instabilities lead to the development of small scale dissipative structures, in which magnetic energy is converted in other forms. In this paper we present three-dimensional numerical simulations of the instability evolution in highly magnetized plasma columns, considering different kinds of equilibria. In fact, the hoop stresses related to the azimuthal component of magnetic field can be balanced either  by the magnetic pressure gradient (force-free equilibria, FF) or by the thermal pressure gradient (pressure-balanced equilibria, PB) or by a combination of the two. FF equilibria are prone to current-driven instabilities (CDI), while PB equilibria are prone to pressure-driven instabilities (PDI).  We perform a global linear stability analysis, from which we derive the different instability properties in the two regimes, showing that PDI have larger growth rates and are also unstable for high wavenumbers. The numerical simulations of the non-linear instability evolution show similar phases of evolution in which the formation of strong current sheets is followed by a turbulent quasi steady-state. PDI  are however characterized by a faster evolution, by the formation of smaller scale dissipative structures and larger magnetic energy dissipation. 

\end{abstract}

\begin{keywords} magnetic reconnection,plasma instabilities
\end{keywords}

\section{Introduction}
The study of current driven kink instabilities in relativistic jets has recently gained a lot of
attention because of the  possible role in the dissipation of the jet magnetic energy that results in the energization of relativistic particles emitting high energy radiation. In fact, the evolution of these instabilities may lead to the formation of reconnection layers where particle acceleration can be very efficient, as shown by PIC simulations \citep{Sironi14, Guo14, Werner17}. Several papers have been devoted to the linear analysis of the instability \citep[see e.g.][]{Bodo13, Bodo19, Sobacchi17, Kim17, Kim18} and to its nonlinear evolution \citep{Mizuno09, Bromberg19, Bodo22} through relativistic magnetohydrodynamic (RMHD) simulations. RMHD simulations have been also performed in order to study the properties of the reconnection layers \citep{Medina21, Kadowaki21} and the observational signatures, by using simplified models for the  accelerated emitting particles \citep{Zhang16, Bodo21}. Particle acceleration resulting from the instability evolution has been recently also studied using PIC simulations \citep{Davelaar20}.

Most of these  studies have  focused on force-free equilibrium configurations. This was motivated by the fact that relativistic jets,  originating in the proximity of a central compact object (in most cases a black hole)  through an interplay of rotation and magnetic processes \citep[see e.g.][]{Komissarov07}, are predicted to be Poynting dominated. 
However, at larger distances,  the longitudinal component of magnetic field decays faster than the azimuthal one, and other forces, like the thermal pressure gradient, may come into play to determine the equilibrium jet configuration. In the case of relativistic jets, only few studies have considered these more general equilibria \citep[see e.g.][]{Oneill12, Alves18, Ortuno22}. In particular, \citet{Oneill12} performed RMHD simulations comparing different kinds of equilibria in which either the pressure gradient or rotation act to balance the hoop stresses of the azimuthal magnetic field component. When the outward pressure gradient contributes to the force balance,  in addition to current driven instabilities, also pressure driven instabilities take place \citep{Kruskal54, Kadomtsev66, Freidberg82, Kersale00, Longaretti08}, which belong to the large class of interchange instabilities.

In \citet{Bodo22} (hereinafter Paper I), we examined the evolution of current driven kink instabilities, comparing several different force-free equilibria and focusing on the dissipation properties. In this paper, we extend that analysis to cases where the pressure gradient contributes to the force balance and to understand, for these cases, the interplay of current and pressure driven instabilities. As a preliminary step, we also present the results of a linear analysis for these more general equilibria. 

The paper is organized as follows: In Section \ref{sec:problem} we introduce the relevant equations and describe the equilibrium configuration, in Section \ref{sec:linear} we do the linear analysis and present its results, in Section \ref{sec:setup} we describe the numerical setup, in Section \ref{sec:results} we present the results of the numerical simulations and, finally, in Section \ref{sec:summary} we give a summary of our findings.

\section{Problem description}
\label{sec:problem}
Our aim is to study instabilities in a highly magnetized, relativistic, plasma column. The governing equations are the equations of relativistic MHD:
\begin{equation}
 \partial_t \left( \gamma \rho \right) + \nabla \cdot \left(\gamma \rho \vec{v}\right)= 0,
  \label{eq:drho/dt} \\ 
\end{equation}
\begin{equation}
\partial_t \left( \gamma^2 w \vec {v} + \vec {E} \times \vec {B} \right) + 
\nonumber
\end{equation}
\begin{equation}
\nabla \cdot 
\left( \gamma^2 w \ \vec {v} \vec {v} - \vec {E} \vec {E} - \vec {B} \vec {B} + (p+u_{\mathrm {em}})\vec{I} \right)= 0,
\label{eq:dm/dt} \\ 
\end{equation}
\begin{equation}
 \partial_t \left( \gamma^2 w -p + u_{\mathrm {em}} \right) + \nabla \cdot \left( \gamma^2 w \vec {v} + \vec {E} \times \vec {B} \right)= 0,
\label{eq:energy}
\end{equation}
\begin{equation} \label{eq:dB/dt}
\partial_t \vec {B} + \nabla \times \vec {E}=0
\end{equation}
where $\rho$ is the proper density,  $p$ is the thermal pressure, $w $ is the relativistic enthalpy $\gamma$ is the Lorentz factor,  $\vec {v}$, $\vec {B}$, $\vec{E}$  are, respectively, the velocity, magnetic field and electric field 3-vectors, $u_{em} = (E^2+B^2)/2$ is the electromagnetic energy density, $\mathbf I$ is the unit $3 \times 3$ tensor and the electric field is provided by the ideal condition  $\vec{E}+\vec{v}\times \vec{B}=0$. In addition, we have to specify an equation of state relating $w$, $\rho$ and $p$,  for which we consider a $\Gamma$-law with constant $\Gamma=5/3$. As it is shown below and also discussed in Paper I, the choice of the equation of state has a little impact on the results. The units are chosen such that the light speed is $c = 1$ and the factor $\sqrt{4 \pi}$ has been absorbed in the fields $\vec{E}$ and $\vec{B}$.   

\subsection{Equilibrium configuration}

We consider an axisymmetric plasma column in the cylindrical coordinates $(r,\varphi,z)$, which is uniform in the azimuthal $\varphi$- and longitudinal $z$-coordinates. It has zero velocity, constant density $\rho_0$ and  a radially varying magnetic field consisting of azimuthal, $B_\varphi$, and longitudinal, $B_z$, components, i.e., $\vec{B} = \left( 0, B_\varphi (r), B_z(r) \right)$. The equilibrium condition in the absence of velocity reads

\begin{equation}
    \dfrac{dp}{dr} = -\dfrac{1}{2r^2}\dfrac{d}{dr}(r^2B^2_\varphi) - \dfrac{1}{2}\dfrac{d}{dr}B^2_z,
\label{eq:radial_balance}
\end{equation}

Eq. (\ref{eq:radial_balance}) leaves the freedom of choosing the radial profile of all the flow variables except one, which is determined by the previous ODE.
In  this work we will  consider the profiles  introduced  by \citet{Bodo13},  where the azimuthal and longitudinal components of magnetic field are given, respectively, by
\begin{equation}
    B^2_{\varphi}(r) = B_0^2\dfrac{a^2}{r^2}\left[1 - \exp{\left(-\dfrac{r^4}{a^4}\right)}\right],
\label{eq:Bphi}
\end{equation}
and 
\begin{equation}
    B^2_z(r) = \dfrac{B_0^2 P^2_c}{a^2} - (1 - \chi)B_0^2\sqrt{\pi}\mathrm{\,erf}\left(\dfrac{r^2}{a^2}\right).
\label{eq:Bz}
\end{equation}

From Eq. (\ref{eq:radial_balance}) we can then derive the profile of the thermal pressure as
\begin{equation}
    p(r) = p_a + \chi\dfrac{B^2_0\sqrt{\pi}}{2}\left[1 - \mathrm{erf} \left(\dfrac{r^2}{a^2}\right)\right],
\label{eq:pressure}
\end{equation}
where $B_0$ characterizes the strength of the magnetic field, $p_a$ is the value of the pressure at large radii, $\mathrm{erf()}$ is the error function and $a$  is the magnetization radius, i.e. the radius inside which most of the magnetic energy is concentrated. 
The above equilibrium depends on four parameters: the pitch on the central axis defined as
\begin{equation}
 P_c = \displaystyle  \lim_{r\to0} \left| \dfrac{rB_z}{B_\varphi} \right|   \,,
\end{equation}
the average jet  cold magnetization, from which we can determine the value of $B_0$, is defined as
\begin{equation}
    \sigma = \frac{\langle B^2 \rangle}{\rho_0  c^2} \,,
\end{equation}
 where the radially averaged magnetic field magnitude squared $\langle B^2 \rangle$ is given by 
\begin{equation}
    \langle B^2 \rangle = \frac{\int_0^a (B^2_z + B_\varphi^2) r dr}{\int_0^a r dr}
\end{equation}
 In addition,  we have the parameter 
 \begin{equation}
     \mathcal{P}_a = \frac{p_a}{\rho_0 c^2},
 \end{equation}
 that defines the ambient pressure in terms of $\rho_0 c^2$ and, finally, the parameter $\chi \in [0,1]$ that allows us to switch from a pure force-free (FF) case ($\chi = 0$), in which the magnetic tension associated to the azimuthal magnetic field component is completely balanced by the gradient of the magnetic pressure associated to the longitudinal field component, to a pure pressure balanced (PB) one ($\chi=1$), where the magnetic tension is instead completely balanced by the thermal pressure gradient. 
 In the first case, the thermal pressure is constant while, in the second case $B_z$ is constant.  
 For intermediate values of $\chi$,  we have hybrid  equilibrium configurations. The PB equilibrium is referred as Z-pinch in the plasma physics literature, while the others are known as screw-pinch.
 \begin{figure*}
    \centering
    \includegraphics[width=0.75\columnwidth]{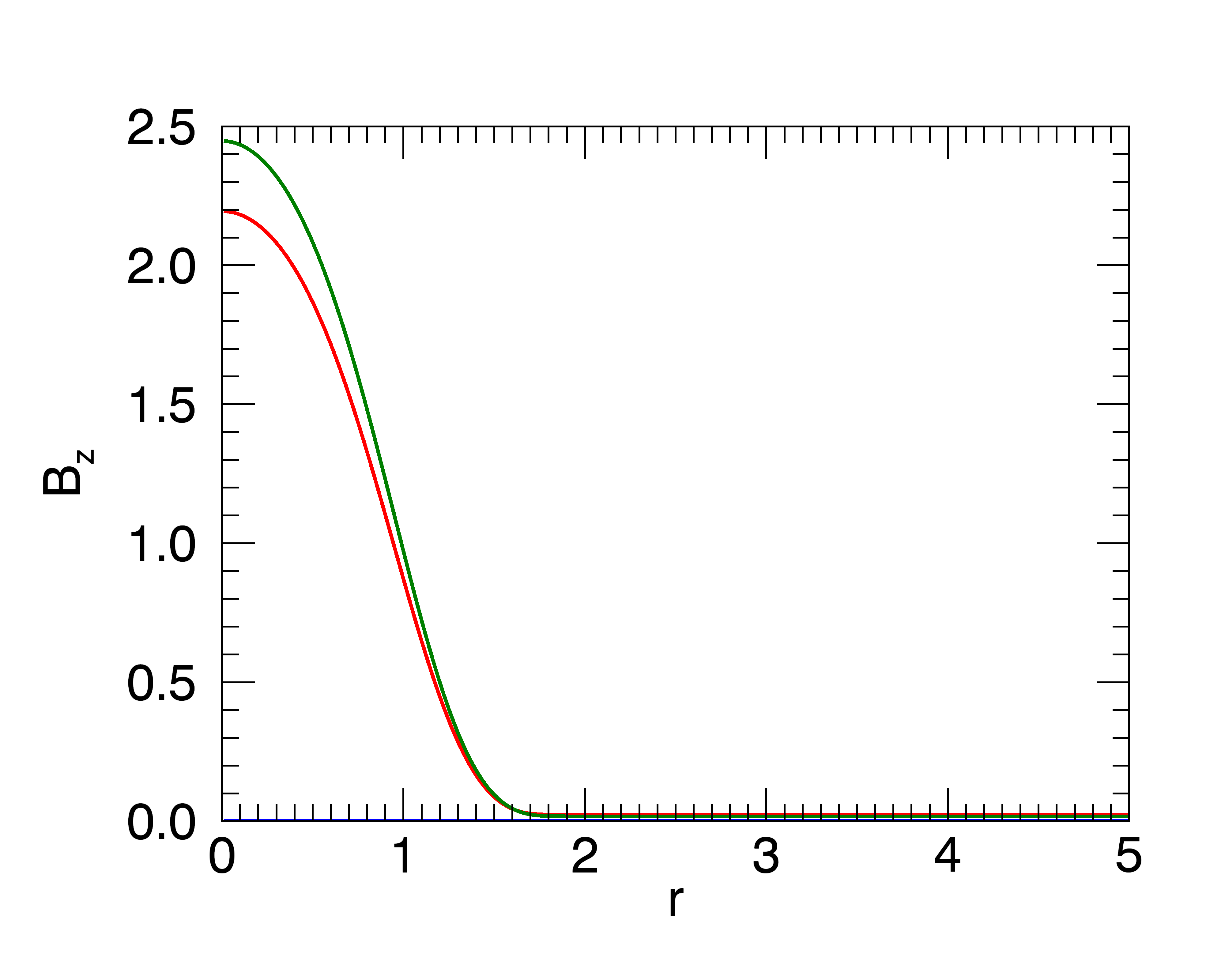}
    \includegraphics[width=0.75\columnwidth]{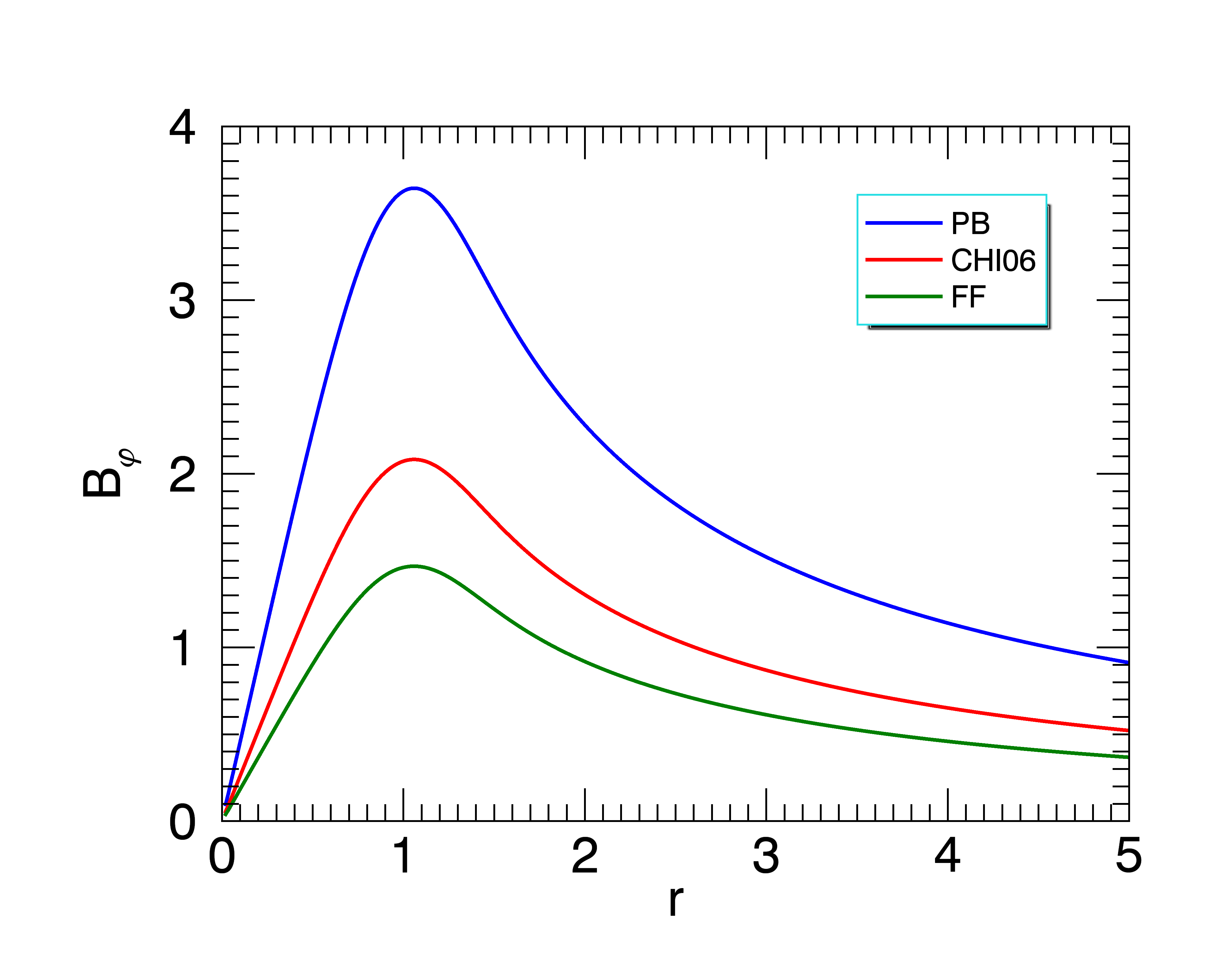} \\
    \includegraphics[width=0.75\columnwidth]{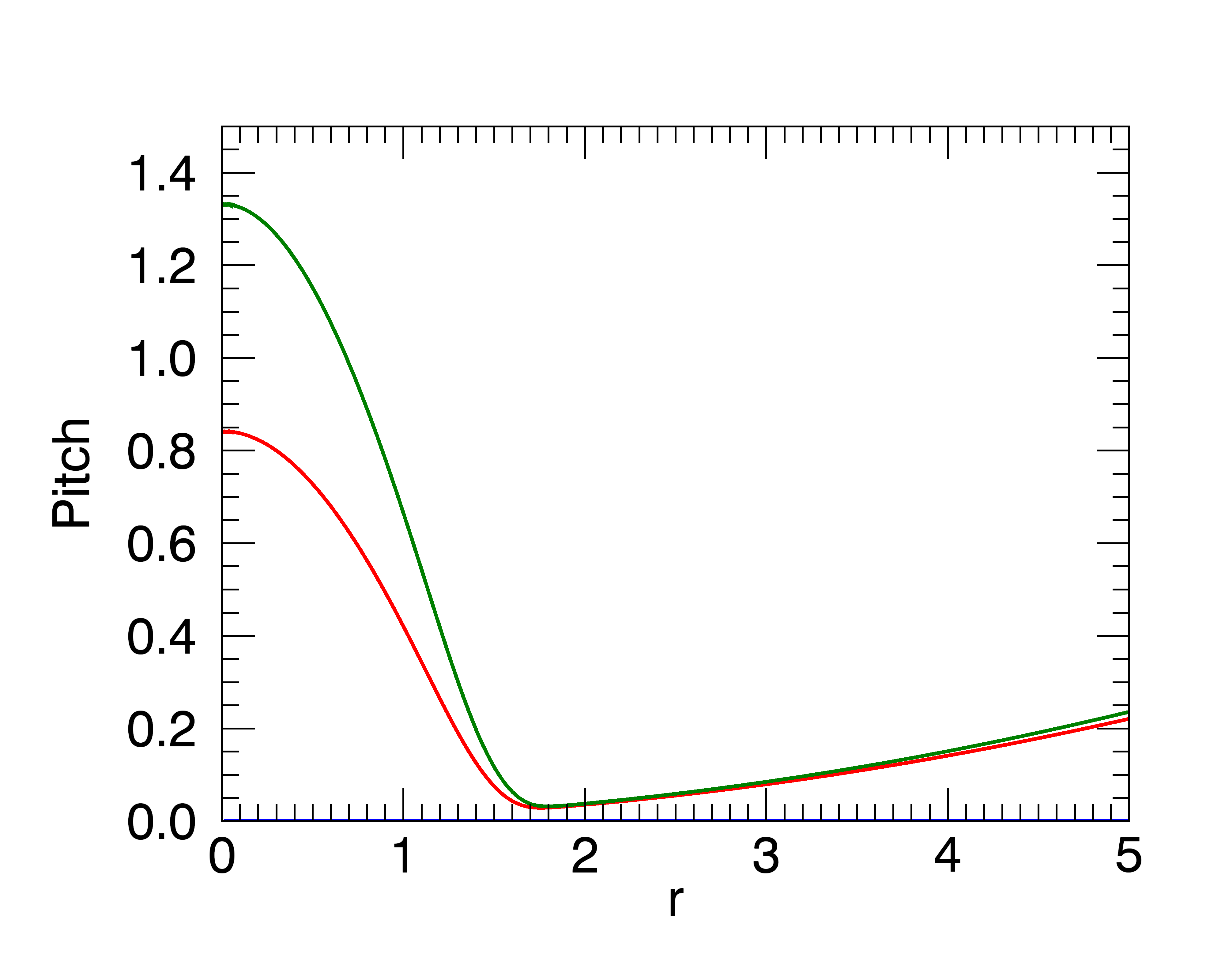}
    \includegraphics[width=0.75\columnwidth]{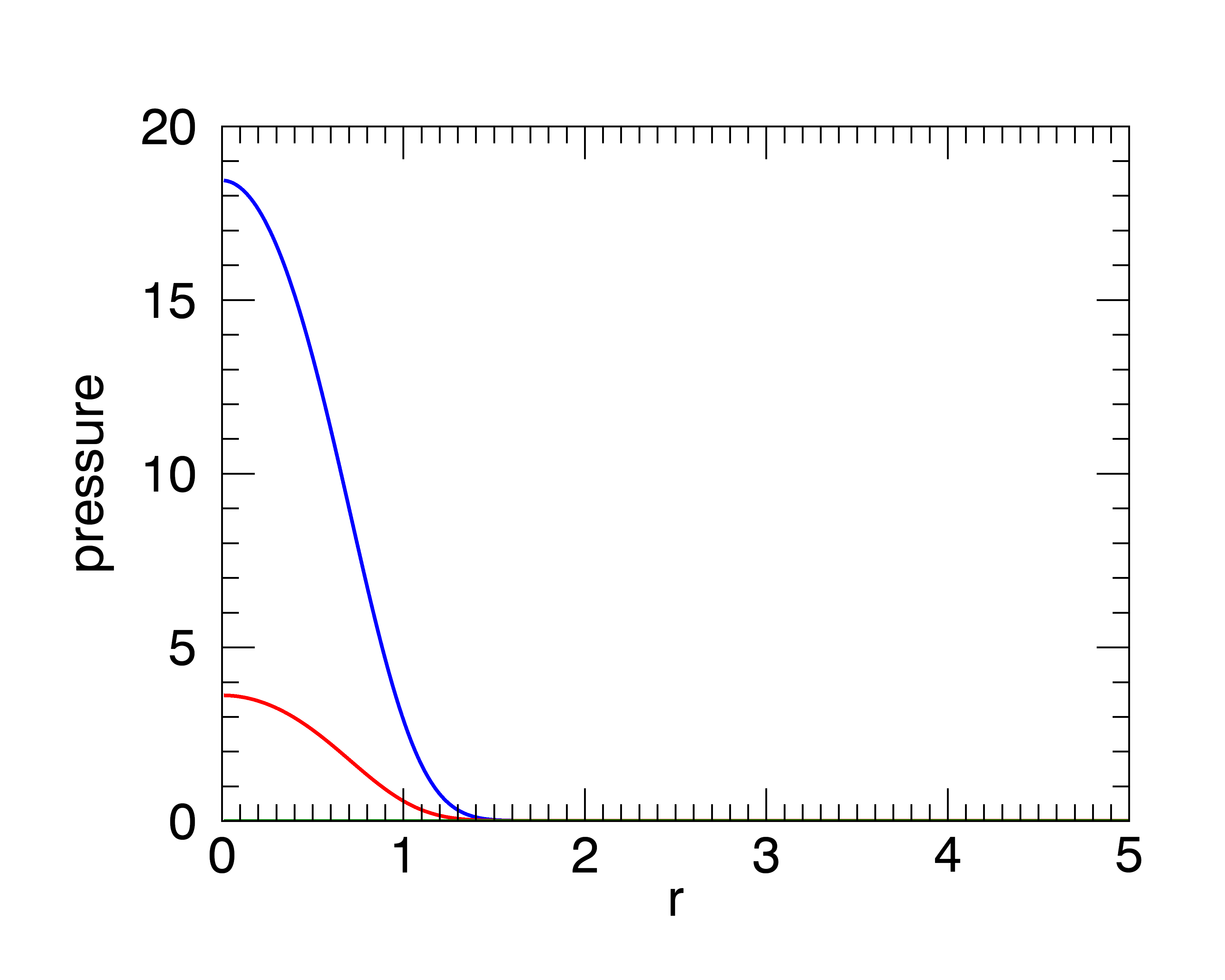}
    \caption{Plots of the initial radial profiles of the longitudinal component of the magnetic field $B_z$ (top left panel), of the azimuthal component $B_{\varphi}$ (top right), of the pitch  (bottom left) and of the thermal pressure $p$ (bottom right). The green, red and blue curves refer, respectively, to the FF, CHI06 and PB equilibria. In the PB case, $B_z=0$ and $P_c=0$, so blue curves are absent in the left top and bottom panels.}
    \label{fig:profiles}
\end{figure*}
 In the following, we will use the magnetization radius $a$ as the unit of length (henceforth putting $a=1$), the initial uniform density $\rho_0$ as the unit of density and, as mentioned above, the speed of light $c$ as the unit of velocity. As a result, the unit of time is  the light crossing time over the magnetization radius, $a/c$, the unit of the magnetic field is $\sqrt{\rho_0} c$ and  the unit of pressure is $\rho_0 c^2$.

  It follows from Eq. (\ref{eq:Bz}) that for a given $\chi$, there is a minimum value of $P_c$ for which an equilibrium solution is still possible,
 \begin{equation}\label{min_Pc}
    P^2_{c,min} = \left( 1- \chi \right) \sqrt{\pi},
 \end{equation}
 while for lower pitch values than this, $B_z^2$ becomes negative and hence the equilibrium can not exist. 
 In the simulations, we consider three different values of $\chi$, namely $\chi = 0,~0.6,~1$, with $P_c = P_{c,min}$ being equal to $1.33,~0.84,~0$ {\bf (for $a=1$)}, respectively.  In Fig. \ref{fig:profiles} we show for  these three cases with the same {\bf cold} magnetization $\sigma = 10$, the  radial profiles of the two magnetic field components, the pitch and thermal pressure. The green and blue curves correspond to the limiting cases FF ($\chi=0$) and PB ($\chi=1$), while the red curve represents the intermediate case ($\chi = 0.6$). In the FF case, the pressure is constant $p_a$, while in the PB case $B_z = 0$. In the intermediate case the equilibrium is maintained by a combination of the gradients of thermal and magnetic pressure.

\section{Linear analysis}
\label{sec:linear}
\begin{figure*}

\includegraphics[width=8.5cm]{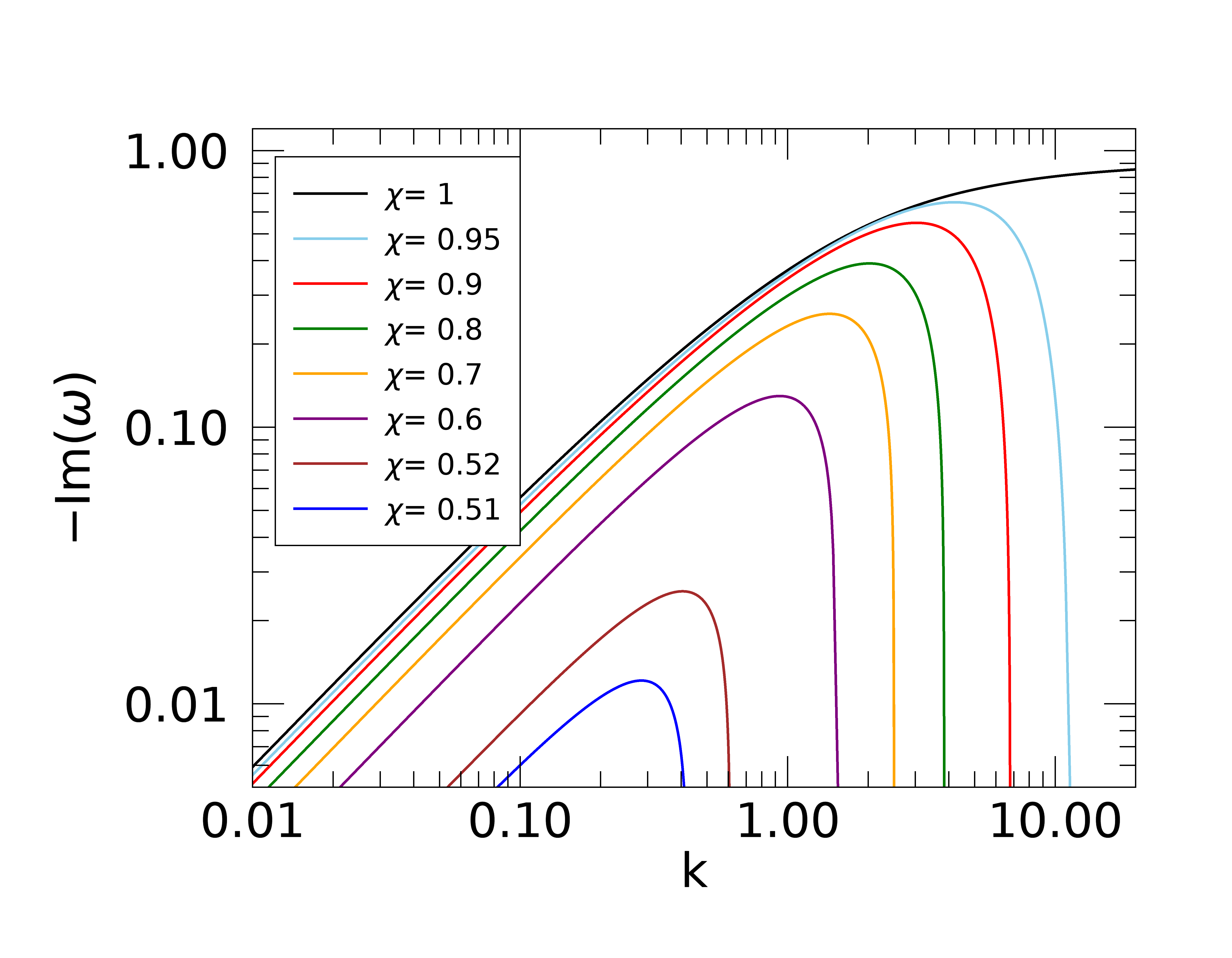}   \includegraphics[width=8.5cm]{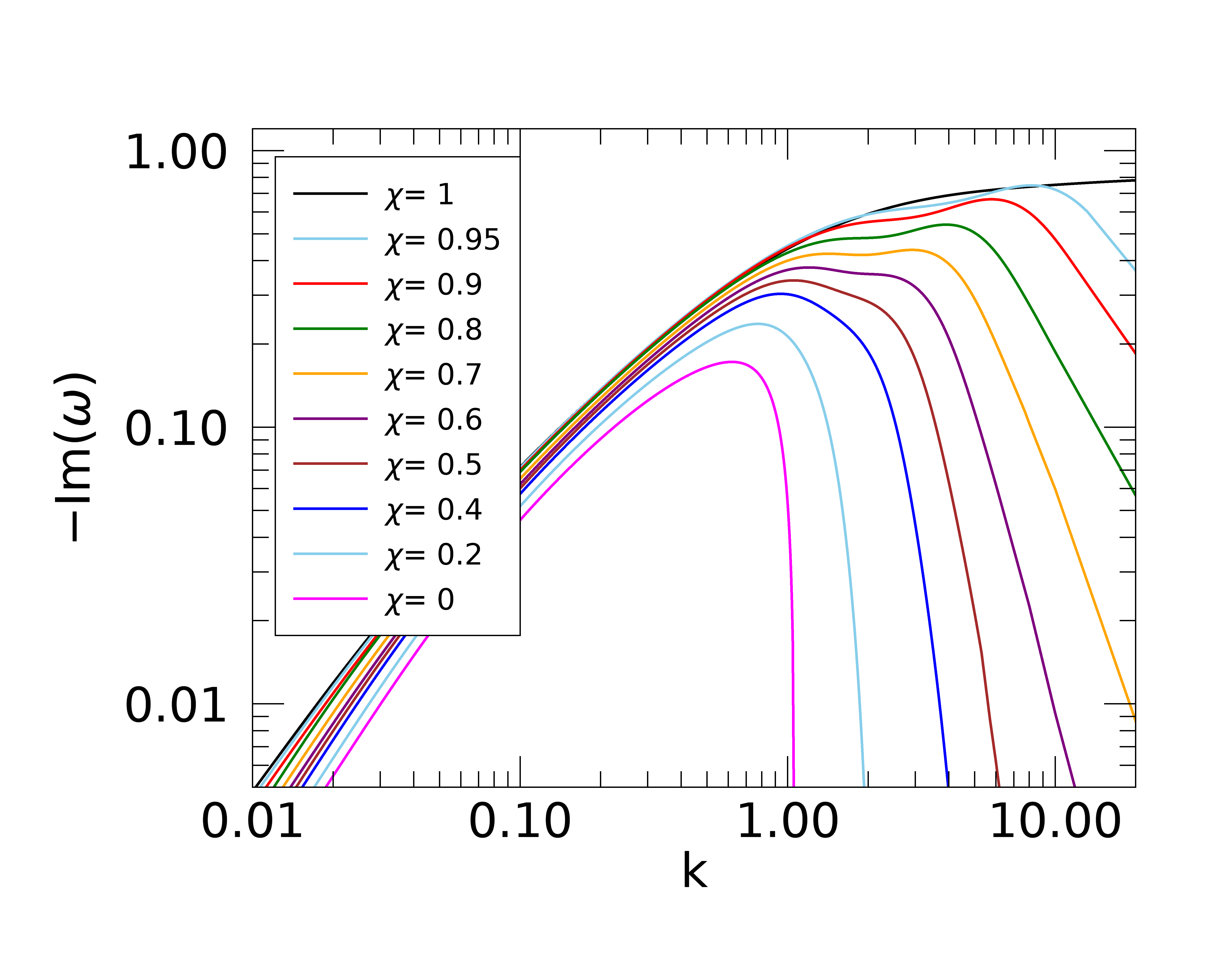}   
\caption{Growth rate, $-{\rm Im}(\omega)$, as a function of the longitudinal wavenumber $k$ for different values of $\chi\in [0, 1]$ for axisymmetric $m=0$ (left) and non-axisymmetric $m=1$ kink (right) modes. With increasing $\chi$, a pure CDI at $\chi=0$ gradually transforms into PDI at $\chi=1$.}\label{fig:linear_results}
\end{figure*}

\begin{figure}
\includegraphics[width=\columnwidth]{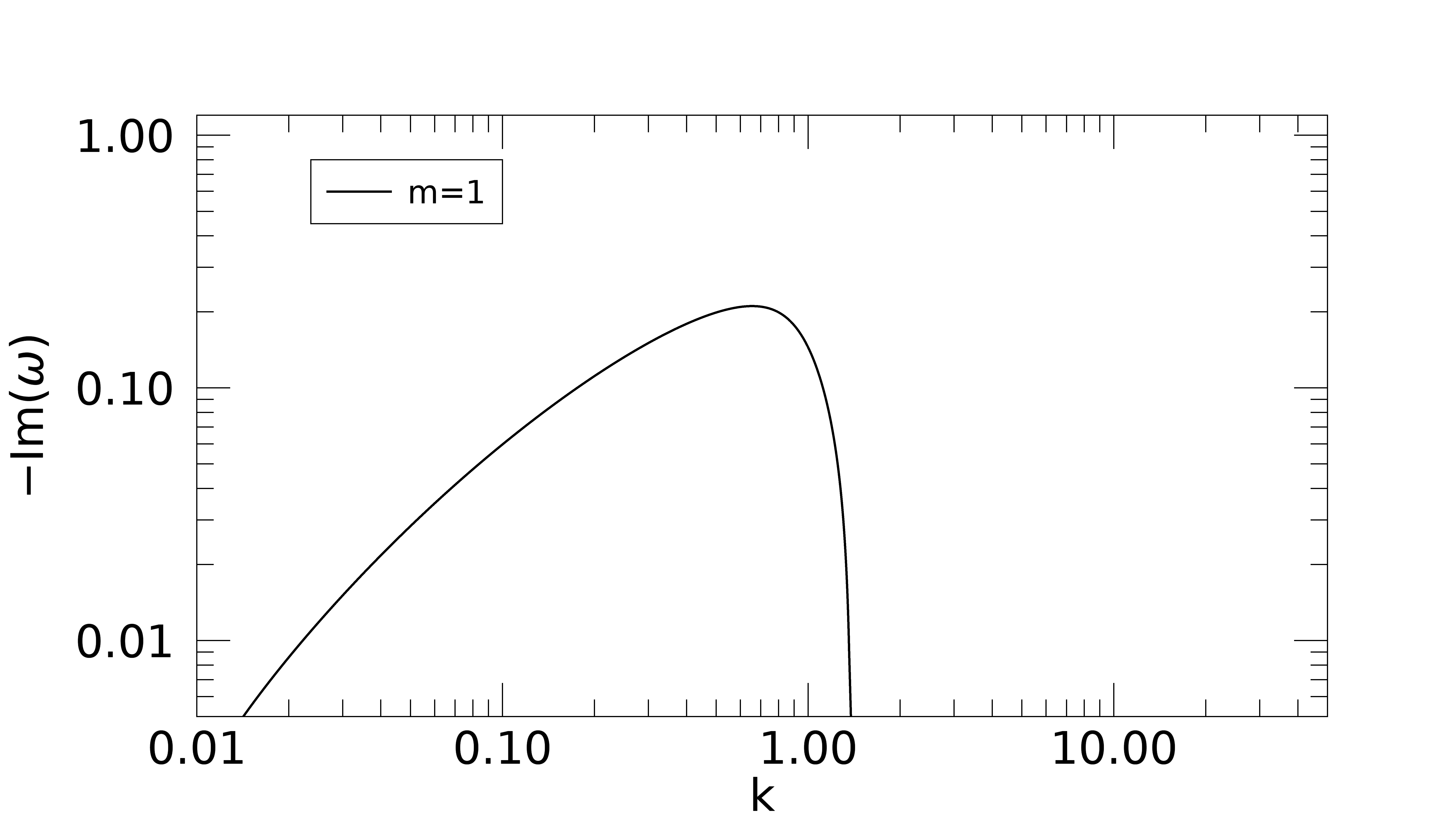}   
\includegraphics[width=\columnwidth]{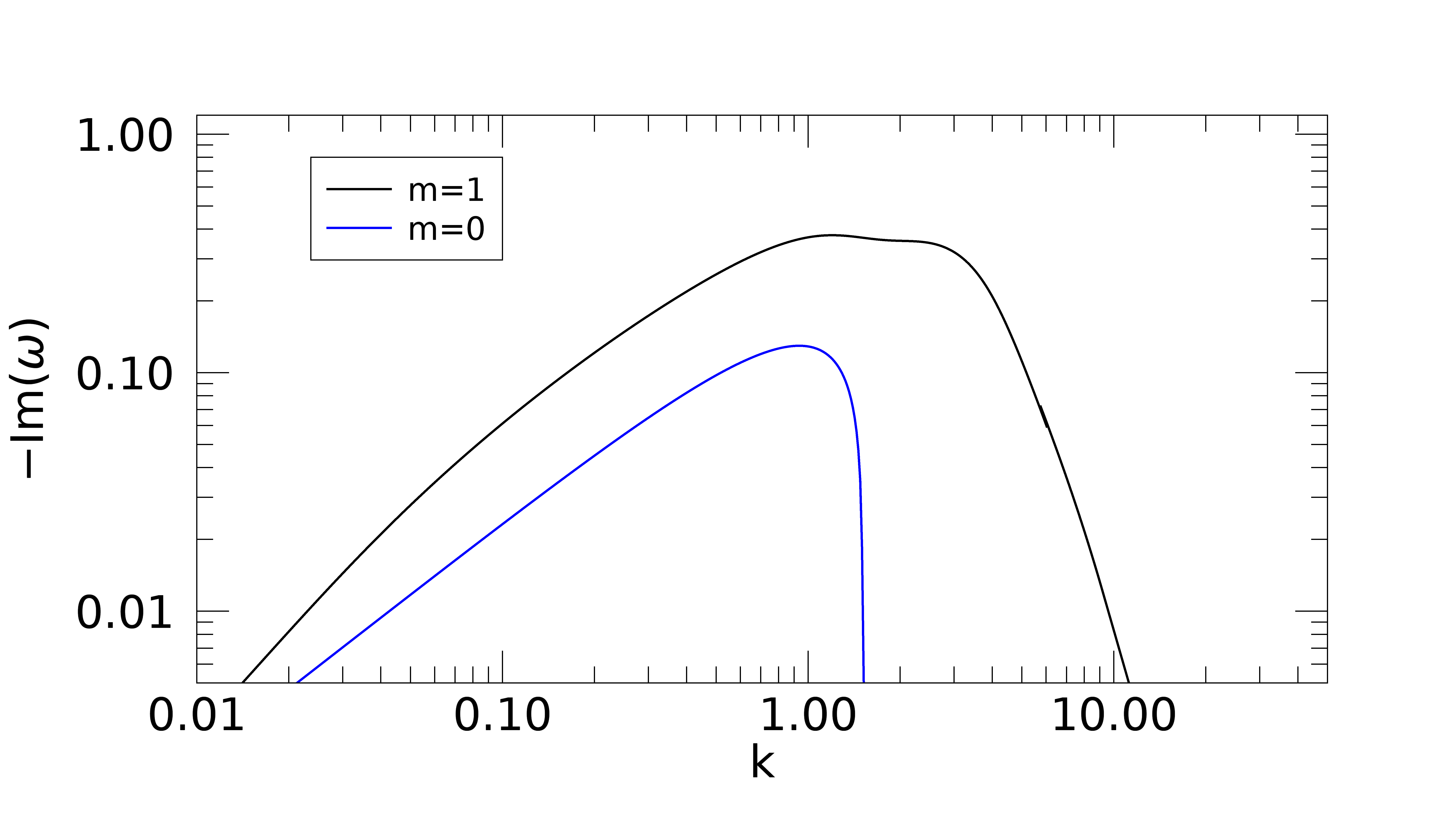}   
\includegraphics[width=\columnwidth]{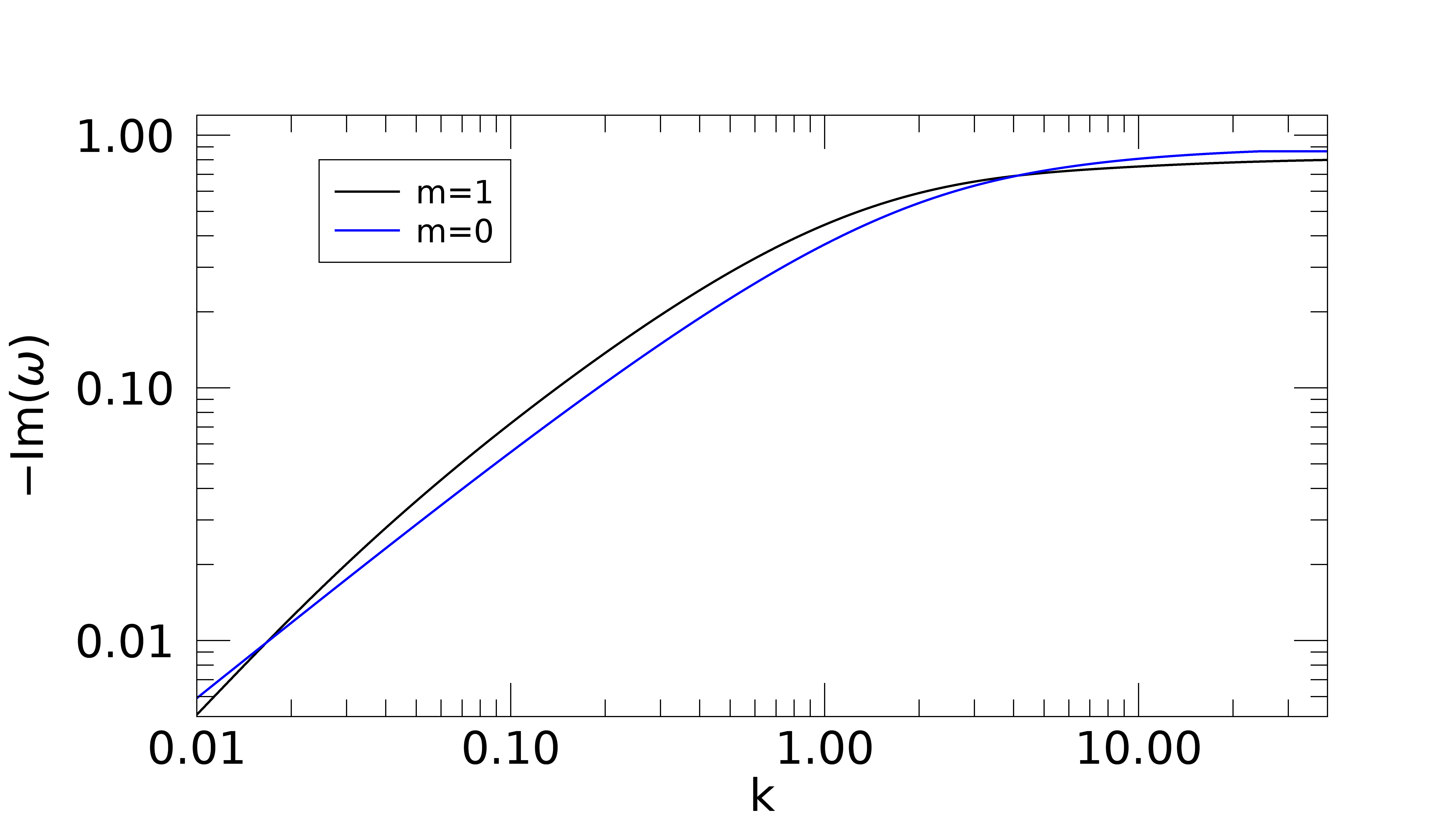}   
 \caption{Plot of the growth rate of $m=0$ (blue curves) and $m=1$ (black curves) modes as a function of $k$ in the FF ($\chi=0$, top),  CHI06 ($\chi=0.6$, middle) and PB ($\chi=1$, bottom) cases. The $m=0$ mode is stable in the FF case (hence not shown), whereas it has the similar growth rate as the $m=1$ mode in the PB case.}
 \label{fig:linear2}
\end{figure}

Before analyzing the simulation results, it is insightful to perform a linear stability analysis of the system. Linear analyses of plasma column configurations in which pressure-driven instabilities are present have been performed, in a local approximation, in the relativistic case by \citet{Begelman98} for a static configuration  and by \citet{NalewaJko12} in the presence of a longitudinal velocity shear. 
More recently \citet{Begelman19} performed a global analysis  for a non-relativistic configuration. 
Here we present a global analysis in the relativistic regime. 

We consider small perturbations $\rho_1$, $p_1$, $\mathbf{v}_1$, $\mathbf{B}_1$ to the equilibrium state defined by Eqs. (\ref{eq:Bphi})-(\ref{eq:pressure}) and linearize the system (\ref{eq:drho/dt})-(\ref{eq:dB/dt}) by assuming the perturbations to be of the modal form $\propto \exp(i\omega t -i m \varphi -i k z)$, where $\omega$ is the complex frequency and $m$ and $k$ are, respectively, the azimuthal and longitudinal wavenumbers. 
After lengthy algebraic manipulations, we arrive at a system of two first order ordinary differential equations in the radial coordinate for the radial displacement $\xi_{1r}=-iv_{1r}/\omega$ and the perturbed total pressure $\Pi_1=\mathbf{B} \cdot \mathbf{B}_1 + p_1$, which can then be written as
\begin{equation}\label{eq:linear1}
    \frac{d \xi_{1r}}{dr} = A_{11} \xi_{1r} + A_{12} \Pi_1 \,,
\end{equation}
\begin{equation}\label{eq:linear2}
    \frac{d \Pi_1}{ dr} = A_{21} \xi_{1r} + A_{22} \Pi_1 \,,
\end{equation}
where the coefficients $A_{11}$, $A_{12}$, $A_{21}$ and $A_{22}$ are given in Appendix A. 
These two equations, supplemented with appropriate boundary conditions near the jet axis $r=0$ and at large radii $r\rightarrow \infty$ (derived in Appendices B and C), represent an eigenvalue problem for $\omega$. 
In the absence of mean velocity, $\omega^2$ is a real number and hence $\omega$ can be purely real or imaginary, in the last case there is instability. 

Following \cite{Bodo13}, the eigenvalue problem is solved by means of a shooting method, using the complex secant method for finding the root $\omega^2$.
At $r=0$, which is a singular point of Eqs. (\ref{eq:linear1}) and (\ref{eq:linear2}), we impose a regularity condition, while at infinity the solutions have the form of radially outward propagating waves that gradually decay with radius. In particular, at $r=0$, we begin the integration at a small distance from the origin where the solution is obtained through a series expansion of the equations, as described in Appendix B. 
Similarly, we start a backward integration from a sufficiently large radius, where the asymptotic solution is obtained as described in Appendix C and then we match the two numerical solutions at an intermediate radius.

This system can have two instability drivers \citep[see e.g.][]{BoydSanderson, Begelman98, Begelman19}: the current parallel to the magnetic field, that leads to current-driven instabilities (CDI), and the pressure gradient, that leads to pressure driven instabilities (PDI). The relative importance of these two drivers depends on the parameter $\chi$.  In addition, while for CDI primarily the $m=1$ kink mode is dominant, for PDI, the axisymmetric $m=0$ mode can be also relevant and comparable to the $m=1$ mode \citep{Begelman98, Begelman19}. 

Figure \ref{fig:linear_results} shows the behaviour of the growth rate of the $m=0$ (left panel) and $m=1$ (right panel) modes as a function of the longitudinal wavenumber $k$.
In each panel the different curves refer to different values of $\chi$.
Note that for each $\chi$ we choose the minimum possible value of $P_c$ from Eq. (\ref{min_Pc}). 
It is seen in this figure that when $\chi = 1$ (i.e., $P_c = 0$), the growth rate first increases as a function of $k$ and then reaches an almost constant value (plateau). 
In this case, there is only PDI and the values of the growth rate are similar for $m=0$ and $m=1$. 
As $\chi$ is lowered, the growth rate and the limiting value of $k$ above which there is stability gradually decrease.
This decrease is more pronounced for the $m=0$ mode, since, for this mode, the stabilizing effect of the longitudinal field $B_{z}$ is more effective \citep[see e.g.][]{Begelman98}. For $m=0$, there is a minimum value of $\chi \sim 0.5$ below which the mode is stable. By contrast, the $m=1$ mode is unstable for any value of $\chi$ and, in particular, for $\chi = 0$ it becomes a pure CDI.

Figure \ref{fig:linear2} shows the behavior of the growth rate as a function of $k$ for those three cases for which we performed the simulations. From top to bottom, the three panels report the growth rates for $\chi = 0,~0.6,~1$, respectively, at the chosen corresponding minimum  pitch $P_c = 1.33, ~0.84,~0$. The blue curve  corresponds to $m=0$, while the black curve to $m=1$. 
As we expect from the results discussed above, in the top panel at $\chi = 0$, the blue curve is absent, since the $m=0$ mode is stable in the FF case. 
With increasing $\chi$, the maximum growth rate increases and moves towards larger wavelengths, with the $m=0$ mode also becoming more prominent. For $\chi = 0.6$,  the $m=1$ mode is still dominant and the maximum growth rate occurs at $k \sim 2-3$. For $\chi = 1$, both modes have similar growth rates which become almost constant for large values of $k$.

Finally, for a comparison with previous related linear stability analyses of jet columns, we note that the dependence of the growth rate on the longitudinal wavenumber $k$ for the PB case (with $B_z=0$) obtained in our global analysis, as shown in Figs. \ref{fig:linear_results} and \ref{fig:linear2}, is in a good agreement with that given by the radially local dispersion relation of PDI in the presence of a dominant azimuthal magnetic field derived in  
\cite{Begelman98} \citep[see also][]{Begelman19}. In both cases, the growth rate exhibits a similar behaviour: 1. monotonically increases with $k$ at small and intermediate values  and forms a plateau at high $k$, where it reaches a maximum, and 2. this maximum growth rate at the plateau is nearly independent of $m$, which is a typical feature of PDI, contrary to CDI which is more sensitive to $m$. \cite{Begelman19} also performed global linear analysis of PDI, however, for a setup different from ours -- a cylindrical annulus with rigid/impenetrable boundary conditions -- that likely resulted in the high-$k$ cutoff of the PDI growth rate and hence stability at high $k$, similar to that in our CHI06 case, instead of the plateau observed in our PB case.

\section{Numerical setup}
\label{sec:setup}
All the simulations were performed on a Cartesian domain with coordinates (in unit of $a$) in the range $x\in[-50,50]$, $y\in[-50,50]$, $z\in[0,17]$ and resolution $N_x \times N_y \times N_z$ is $768 \times 768 \times 264$, except for one run (PB1) with a lower resolution $384 \times 384 \times 132$. We adopted a uniform grid in the central region $x,y\in[-13,13]$ and $z \in[0,17]$ while a geometrically stretched grid was used in the remaining regions. The length of the domain in the $z$-direction, $L_z=17$, is chosen to be approximately twice the wavelength $\lambda_m=2 \pi/k_m$ of the CDI mode with the maximum growth rate in the FF case, which has the smallest wavenumber $k_m\approx 0.7$ compared to the most unstable wavenumbers in the CHI06 and PB cases, as seen in Fig. \ref{fig:linear2}. 

As done in Paper I,  initially at $t=0$ the equilibrium configuration is perturbed with a small radial velocity of the form
\begin{equation}
    v_r = \varepsilon r \exp{(-r^4)}\sum_{n=1}^{N}\sin\left({\dfrac{2\pi nz}{L_z}+\phi_n}\right),
\end{equation}
where $\varepsilon$ is an amplitude, chosen as $\varepsilon = 0.01$, $N$ is the number of modes with different longitudinal wavenumbers, chosen as $N=25$, and $\phi_n$ are random phases.

We consider three different cases corresponding to three different values of $\chi$, namely $\chi=0$ (FF, this case has already been presented in Paper I), $\chi = 0.6$ (CHI06, hybrid, or mixed case) and $\chi = 1$ (PB). In the three cases, we have different values of $P_c$, which, in each case, is set equal to the corresponding $P_{c, min}$, implying $B_z\rightarrow 0$ at large $r \rightarrow \infty$. In all the cases, we employ the same magnetization parameter $\sigma = 10$, which, as defined above, represents the cold magnetization. If we consider the hot magnetization 
\begin{equation}
    \sigma_h = \frac{\langle B^2 \rangle }{\langle w \rangle}\,,
\end{equation} 
its corresponding values will be different for the different cases, more precisely: $\sigma_h \approx 1.667$ for the PB case, $\sigma_h \approx 2.5$ for the CHI06 case and $\sigma_h \approx 10$ for the FF case. These differences can be equivalently expressed in terms of the standard plasma $\beta$ parameter
\begin{equation}
    \beta = \frac{2 \langle p \rangle }{\langle B^2 \rangle}
\end{equation}
which is thus equal to $\beta \approx 0.38$ for the PB case, $\beta \approx 0.23$ for the CHI06 case and $\beta \approx 0.01$ for the FF case. We also performed two additional simulations: the above-mentioned PB1 case that has half the resolution of the reference case PB and PB2 case that uses the Taub-Matthews equation of state \citep{Mignone05}, which is an approximation of the exact Synge equation of state. Table \ref{tab:setup_summary} summarizes all the simulations performed with the corresponding parameter values.  

To better keep track of the magnetized column evolution,
we augment the RMHD system of equations with the evolution of a passive tracer $f$,
\begin{equation}
    \dfrac{\partial{(\gamma\rho f)}}{\partial{t}} + \nabla \cdot(\gamma\rho f \mathbf{v}) = 0.
    \label{eq:tracer}
\end{equation}
The passive tracer serves as a coloured fluid, allowing us to better study the mixing  between the magnetized column and the external medium. At $t=0$, we set $f$ to $1$ inside the magnetization radius and to $0$ outside. During the evolution, $f$ will take values between 0 and 1 as a result of the mixing process.  

The simulations were performed with the PLUTO code \citep{PLUTO}, using a parabolic reconstruction, HLL Riemann solver and a constrained transport method \citep{Balsara99, Londrillo04} to keep the $\nabla \cdot \mathbf{B} = 0$ condition under control. The boundary conditions for all the simulations are standard outflow conditions in the $x$ and $y$ directions and periodic in the $z$ direction. 

\begin{table*}
    \centering
    \begin{tabular}{c|c|c|c|c|c|l} 
    \hline
    Simulation   &  $\chi$  & $P_c$  & $\sigma$    &  $t_{stop}$ & $N_x \times N_y \times N_z$ & Eq. state  \\
    \hline\hline
    PB & 1 & 0 & 10 & 125 & $768 \times 768 \times 264$ &  $\Gamma$-law \\ 
	\hline 
    PB1 & 1 & 0 & 10 & 125 & $ 384 \times 384 \times 132$ &  $\Gamma$-law \\ 
	\hline 
    PB2 & 1 & 0 & 10 & 125 & $768 \times 768 \times 264$ &  Taub-Matthews \\ 
	\hline 
    CHI06 & 0.6 & 0.842 & 10 & 250 & $768 \times 768 \times 264$ &  $\Gamma$-law  \\ 
	\hline 
	FF & 0 & 1.333 &  10  & 550 & $768 \times 768 \times 264$ &  $\Gamma$-law \\ 
	\hline
    \end{tabular}
    \caption{List of the simulations with the respective parameter. The first column is the name of the simulation, the second column is the value of $\chi$, the third is the value of  $P_c$, the fourth is the value of magnetization  $\sigma$, the fifth is the final time of the simulations, the sixth column is the numerical resolution in the computational box and, finally, the seventh column specifies the equation of state used. }
    \label{tab:setup_summary}
\end{table*}

\section{Results}
\label{sec:results}

\subsection{Instability evolution}

\begin{figure*}

    \includegraphics[width=5cm]{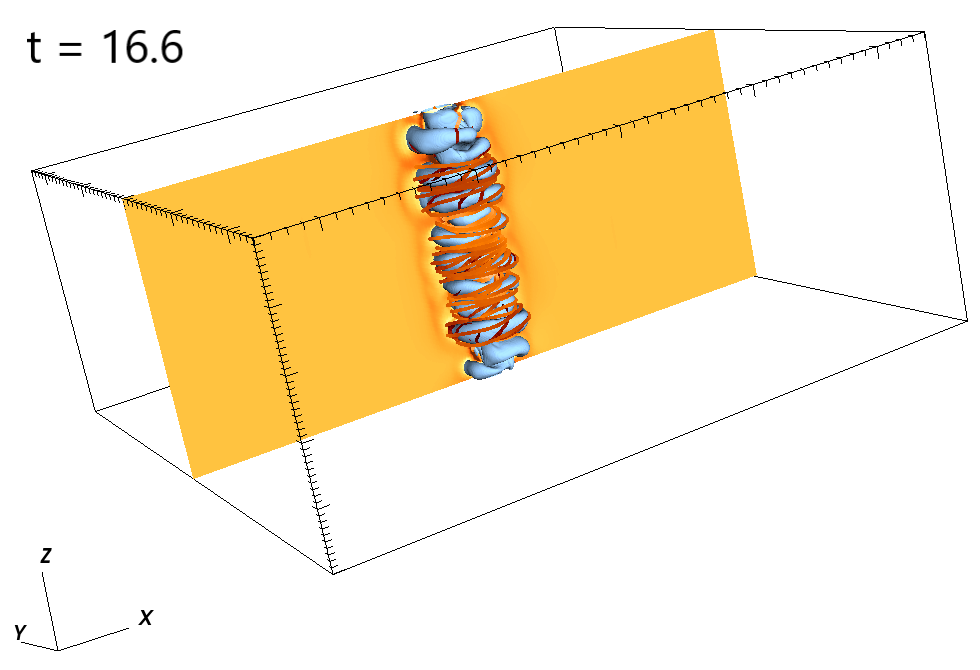}   
    \includegraphics[width=5cm]{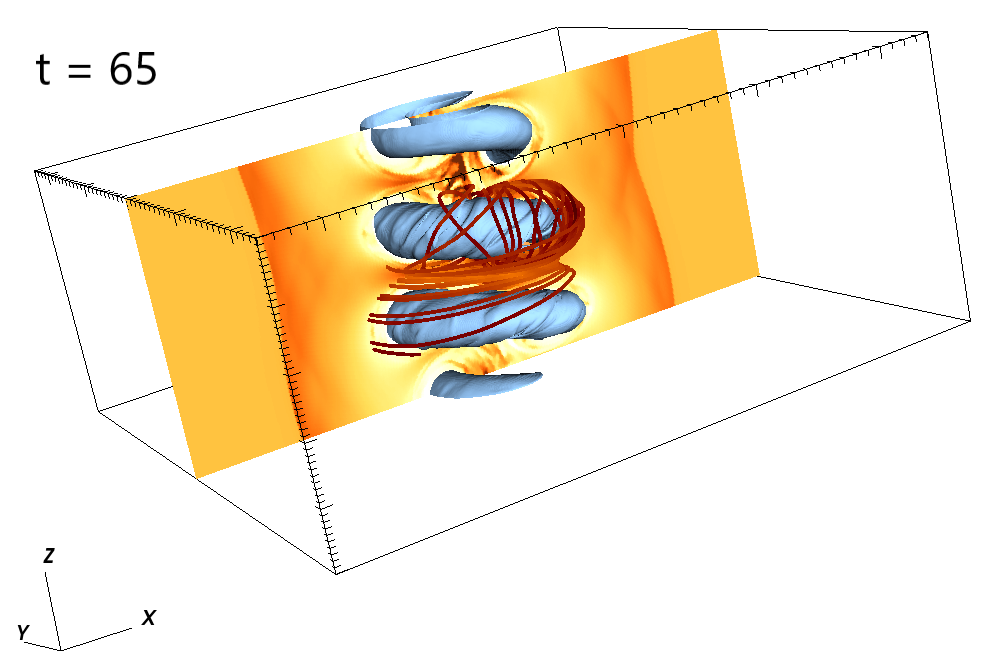}  
     \includegraphics[width=5cm]{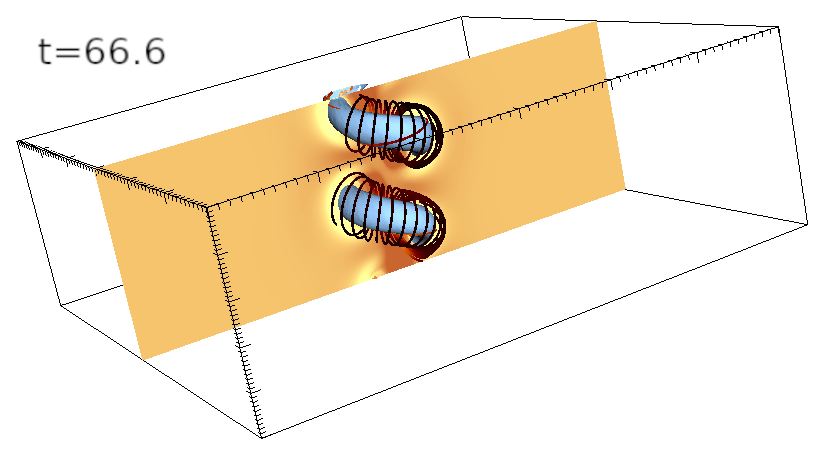}  \\
  \hspace{5mm}
  
    \includegraphics[width=5cm]{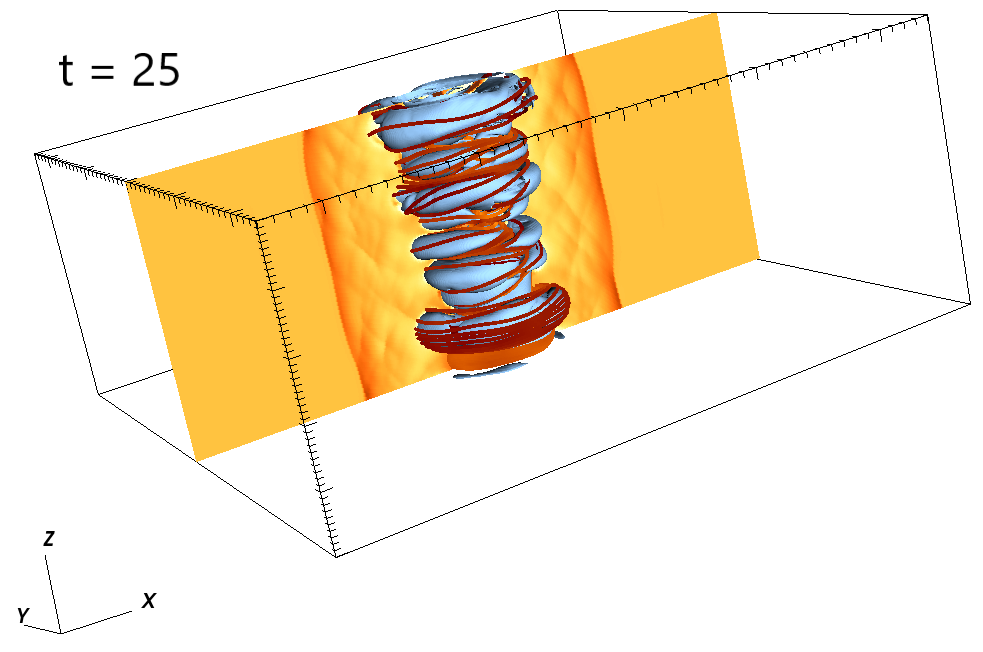} 
    \includegraphics[width=5cm]{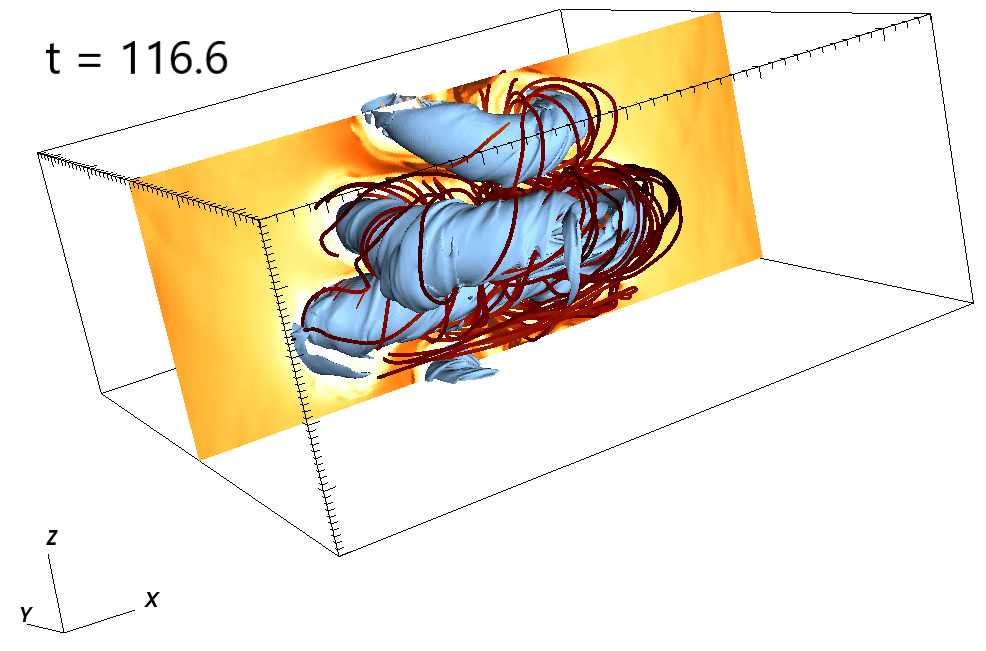}  
    \includegraphics[width=5cm]{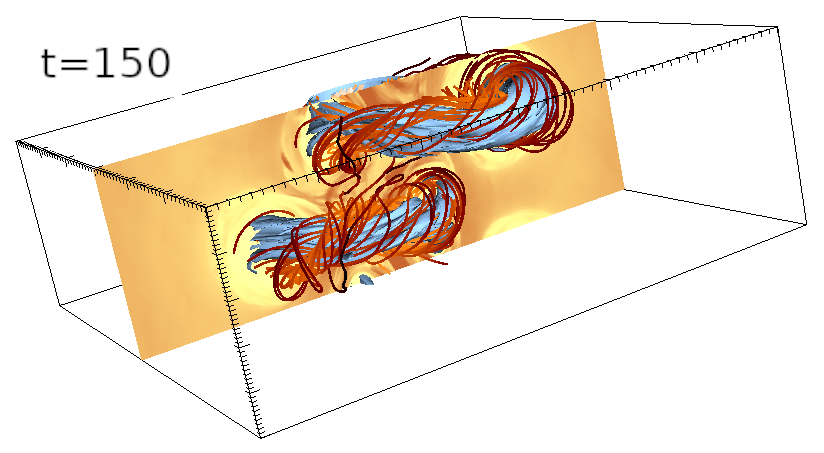}  \\
  
  \hspace{5mm}
 
    \includegraphics[width=5cm]{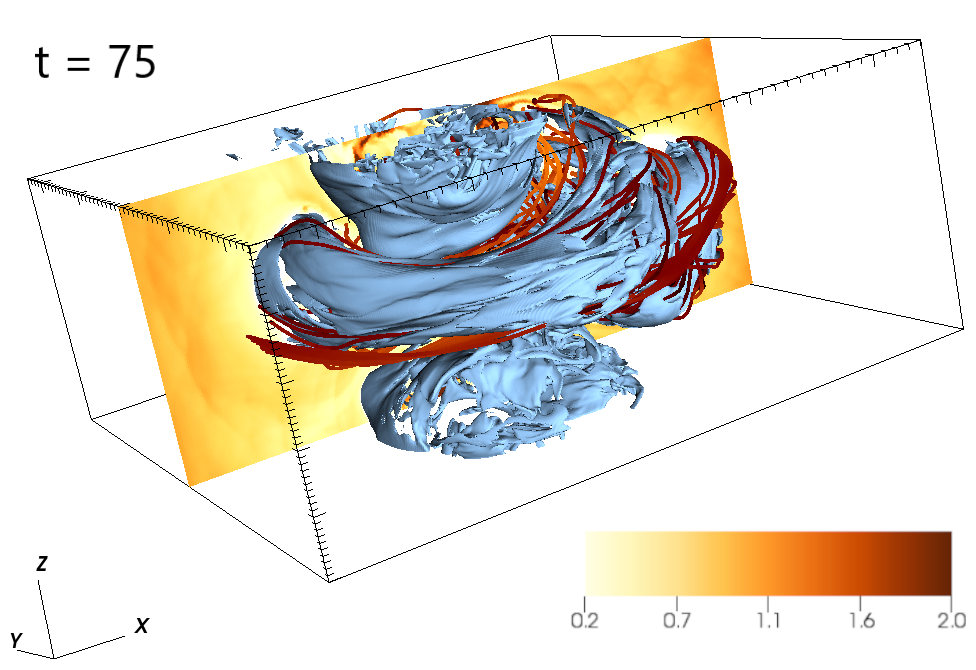}   
    \includegraphics[width=5cm]{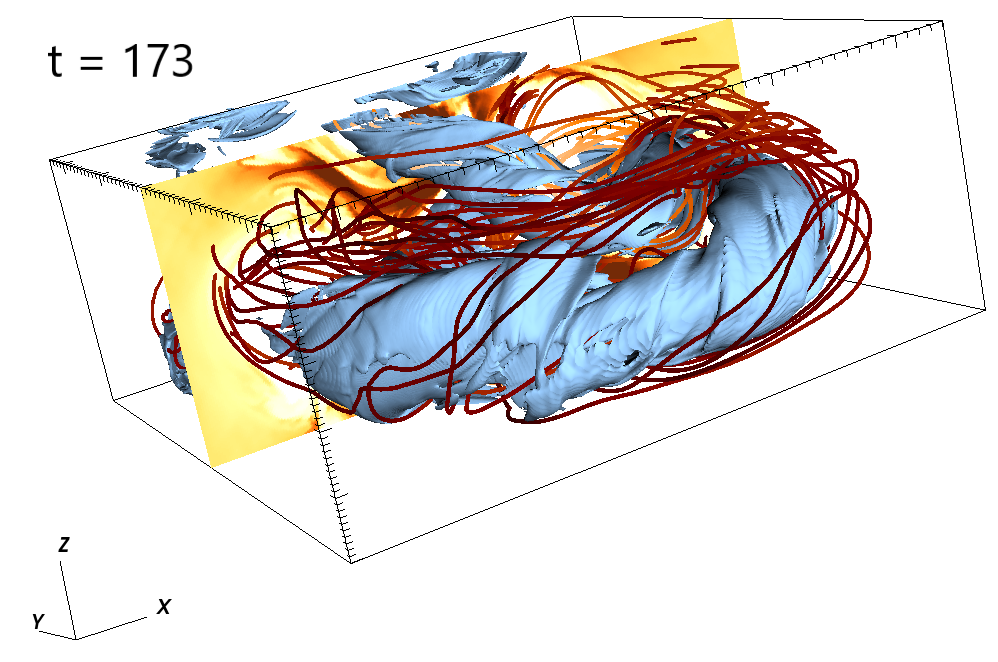}
    \includegraphics[width=5cm]{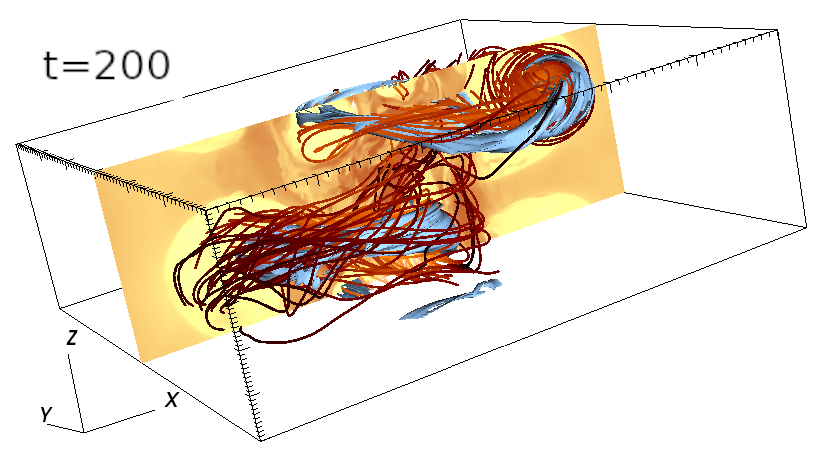}

    \caption{Evolution of the instability from the linear growth to nonlinear saturation and turbulence. In each panel, we display a three-dimensional isocontour of the tracer distribution (in light blue) with representative magnetic field lines in red and a cut of the density distribution in the $x-z$ plane at $y=0$. For all the simulations the limits of the displayed box are $-12\leq x,y \leq 12$ and $0\leq z\leq 10$. The left column displays the evolution for the PB case, the middle one for the hybrid CHI06 case and the right one for the FF case. Snapshots in each row represent a specific stage of the evolution: maximum of the dissipation (top row), end of the peak and beginning of the turbulent phase (middle row) and the final, fully turbulent phase (bottom row). Each stage depicted in a given row occurs, however, at different times in these three cases due to the differences in the corresponding instability growth rates.}
    \label{fig:PB_3D}
\end{figure*}

\begin{figure*}
    \centering
    \includegraphics[width=\textwidth]{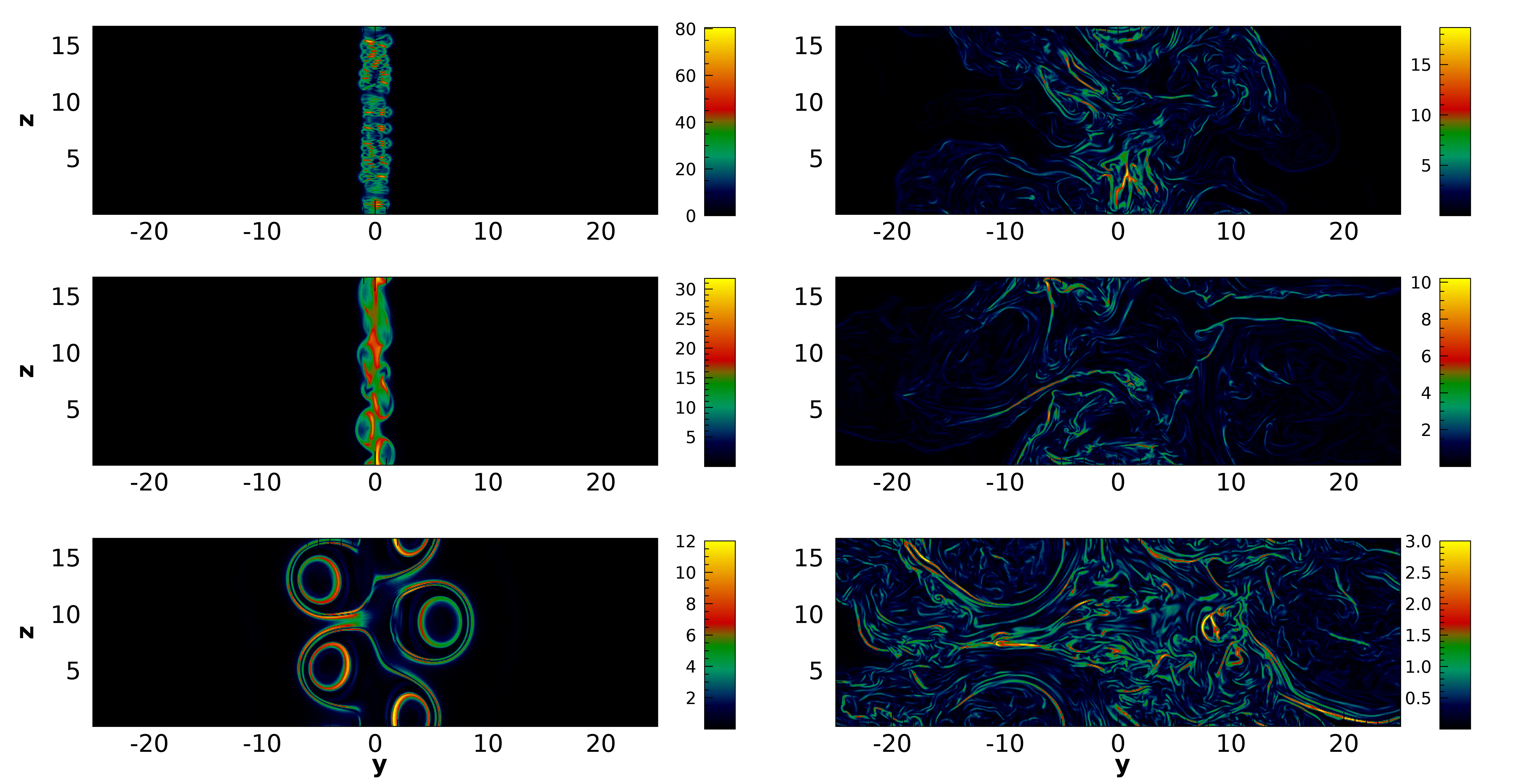}
    \caption{Two-dimensional sections of the current density distribution in the $y-z$ plane at $x=0$ for all the cases, at two different times. The top row refers to the PB case, the middle row to the CHI06 case and the bottom row to the FF case. The panels on the left correspond to a time around the dissipation peak and are, respectively, at $t=11.6$ for the PB case, $t = 70$ for the CHI06 case and $t=80$ for the FF case. The panels on the right correspond to a time when the quasi-steady turbulent phase has already developed and are, respectively, $t=90$ for the PB case, $t=225$ for the CHI06 case and $t=250$ for the FF case.}
    \label{fig:curr}
\end{figure*}

\begin{figure*}
   \includegraphics[width=\textwidth]{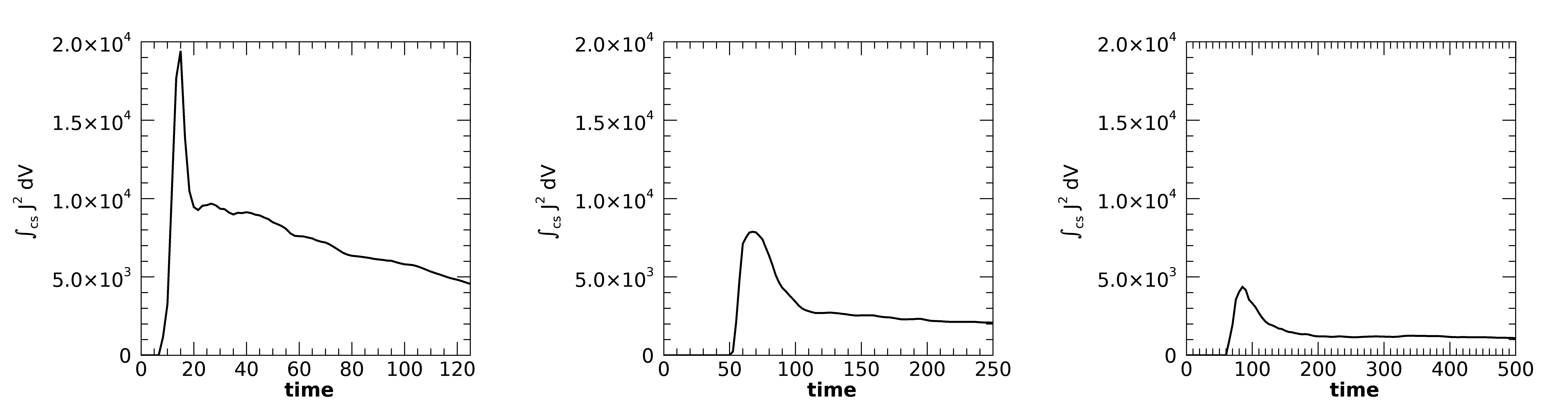}  
    \caption{Evolution of the integral of $j^2$ in the current sheet regions in the PB (left), hybrid CHI06 (middle) and FF (right) cases.}
    
    \label{fig:Jcs}
\end{figure*}

In Fig. \ref{fig:PB_3D} we show the instability evolution for the considered three cases by displaying the 3D composite views of the jet at three different times. Each panel shows an isosurface of the tracer distribution (light blue), a two-dimensional section of the density distribution in the $x-z$ plane and a set of representative magnetic field lines. 
Each column refers to a different case: left column to the PB, the middle column to the hybrid CHI06 and the right column to the FF cases, while each row to a different time. 
As discussed in Paper I, the instability evolution proceeds in two phases: first the helicoidal perturbations grow and saturate, leading to the formation of strong current sheets (see also the left panels of Fig. \ref{fig:curr}), where the magnetic energy is dissipated, corresponding to a peak in the energy dissipation rate (see also Fig. \ref{fig:Jcs}). 
Afterwards, quasi-steady turbulence sets in and the current sheets become more fragmented and irregular (see right panels of Fig. \ref{fig:curr}), continuing to dissipate magnetic energy, but at a smaller rate. In each row of Fig. \ref{fig:PB_3D}, the snapshot times for the instability structures from the different simulations are not the same but chosen such that they correspond to similar key phases of the instability evolution. Specifically, in the top row the time is chosen to be close to the maximum (peak) of the dissipation, in the middle row the time is chosen at the end of the peak and beginning of the turbulent phase, while in the bottom row the time is taken during the fully developed turbulent phase. 

\begin{table}
    \centering
    \begin{tabular}{|c|c|c|c|}
    \hline
        & PB & hybrid & FF \\
    \hline
    top row    &  $ f = 0.1$  &  $ f = 0.1$    &    $f=0.8$ \\
    \hline
    middle row    & $ f = 0.1$  &  $ f = 0.1$ &  $f=0.7$  \\
    \hline
    bottom row    & $ f = 0.05$  & $ f = 0.1$ &  $f=0.5$\\
    \hline 
    \end{tabular}
    \caption{Summary of the tracer $f$ values for each image in Fig. \ref{fig:PB_3D}. Note that due to the different mixing ratios for each case, the displayed value of the tracer is case-dependent. }
    \label{tab:3D_summary}
\end{table}

The instabilities that arise in the FF and PB equilibria have a different physical origin -- current-driven for the first one and pressure-driven for the second one. As discussed in Section \ref{sec:linear}, this is reflected in the different properties of these instabilities, like the maximum growth rate and the the corresponding dominant wavelength. 
Our linear analysis has shown that the dominant PDI modes have shorter wavelengths and larger growth rates than the dominant CDI mode modes.  For PDI, also the $m=0$ mode is unstable and has a growth rate comparable to that of the $m=1$ mode. Specifically, it is seen from Fig. \ref{fig:linear2} that in the PB case, the growth rates for the $m=0$ and $m=1$ modes are comparable and almost constant for $k \gtrsim 10$; in the CHI06 case, the $m=1$ mode is dominant and the growth rate has a flat maximum over the range $1\lesssim k \lesssim 3$ (but still somewhat higher at $k\approx 1.1$) and, finally, in the FF case, the $m=0$ mode is stable and the growth rate of the $m=1$ mode has a peak for $k \approx 0.7$. These results from the linear analysis  are in fact confirmed by the simulations: in the first row  of Fig. \ref{fig:PB_3D} we see that the dominant wavelength of the instability increases as we move from the PB to FF cases. Indeed, in the PB case, we observe a superposition of short wavelength modes, whereas in the CHI06 and FF cases, we observe the larger helicoidal deformations with $m=1$ and clearly noticeable longitudinal wavelengths about $\lambda = L_z/3$ and $\lambda = L_z/2$, respectively. These wavelengths in fact correspond to the wavenumbers $k_m\approx 1.1$ and $k_m\approx 0.7$ of the highest growth rate in the CHI06 and FF cases, respectively, as obtained from the above linear analysis (Fig. \ref{fig:linear2}).  Therefore, we can associate the structures seen in Fig. \ref{fig:PB_3D} for the CHI06 and FF cases with the dominant most unstable modes  (this comparison between the linear analysis and simulation results is further addressed in Appendix D). 

As the instability develops into the nonlinear regime, the size of spatial structures tend to increase as it is shown by the middle and bottom panels of Fig. \ref{fig:PB_3D}. By comparing the times of each snapshot, it is evident that the instability growth rate and the corresponding evolution  are much faster in the PB case than those in the FF one. 
At later times, the contours of the passive tracer $f$ become more corrugated and the magnetic field lines more aligned to the helicoidal shape of the tracer contours.  
The progressively corrugating shapes of the isocontour  surfaces indicate that a mixing process is going on between the magnetized column and the ambient medium, which proceeds with different intensity in these three cases. This is seen in Table \ref{tab:3D_summary}, giving the values of the tracer isocontours at three stages of the evolution illustrated in Fig. \ref{fig:PB_3D}, note how these values are different in the PB, CHI06 and FF cases.

\begin{figure}
    \centering
    \includegraphics[width=0.5\textwidth]{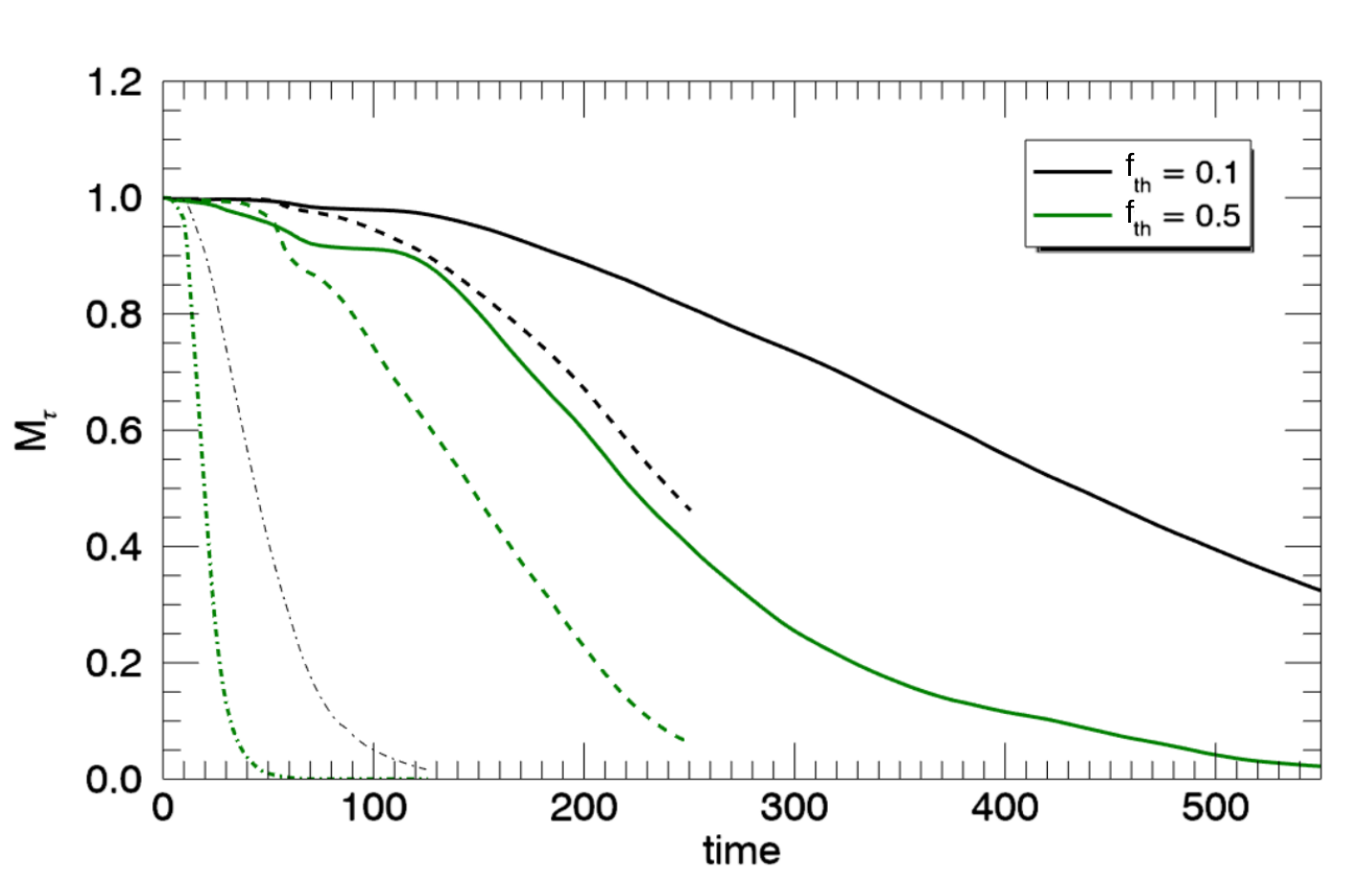}
    \caption{Plot of the tracer mass fraction for $f > f_{th}$. Here we consider two different thresholds for the tracer variable, $f_{th} = 0.1$ (black) and $f_{th} = 0.5$ (green). The solid line refers to the FF case, the dashed one to the CHI06 case and the dot-dashed line to the PB case.}
    \label{fig:Mixing}
\end{figure}

A more quantitative measure of the mixing process is presented in Fig. \ref{fig:Mixing}, showing the fraction $M_{f}(t)/M_{f}(0)$ as a function of time, where $M_{f}$ is the total mass for the tracer $f$ larger than a given threshold $f_{th}$, 
\begin{equation}
    M_{f}(t,f>f_{th}) = \int_{f > f_{th}}\gamma\rho f dV,
    \label{eq:tracer_mass}
\end{equation}
for two different  values of the threshold, $f_{th} = 0.1$ (in black) and $f_{th} = 0.5$ (in green), for all the three cases: FF (solid line), CHI06 (dashed line) and PB (dot-dashed line).
\begin{figure}
  \begin{minipage}{0.47\textwidth}
    \centering
  
    \includegraphics[width=0.89\columnwidth]{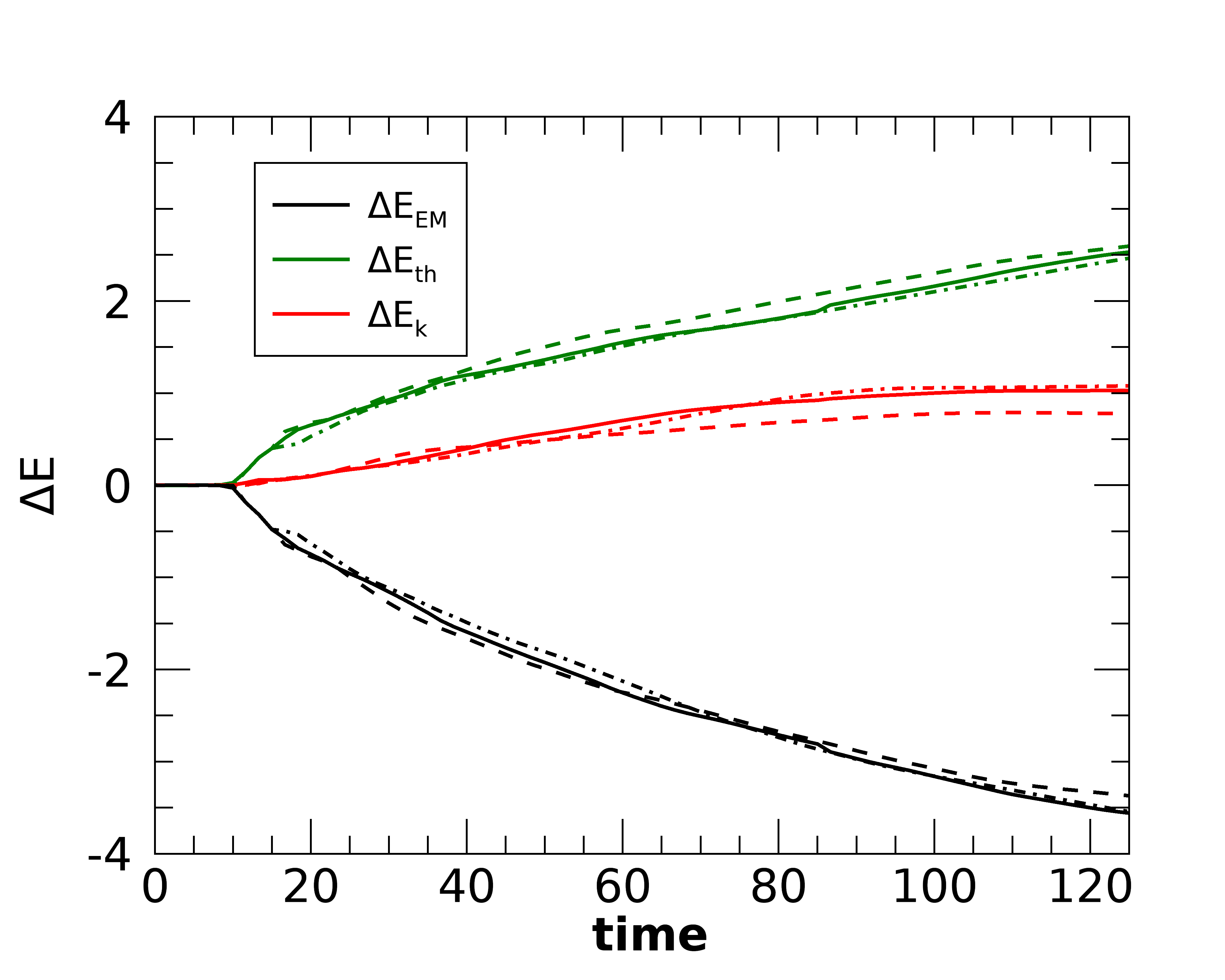}     \end{minipage}
  \hspace{5mm}
  \begin{minipage}{0.47\textwidth}
    \centering  
    \includegraphics[width=0.89\columnwidth]{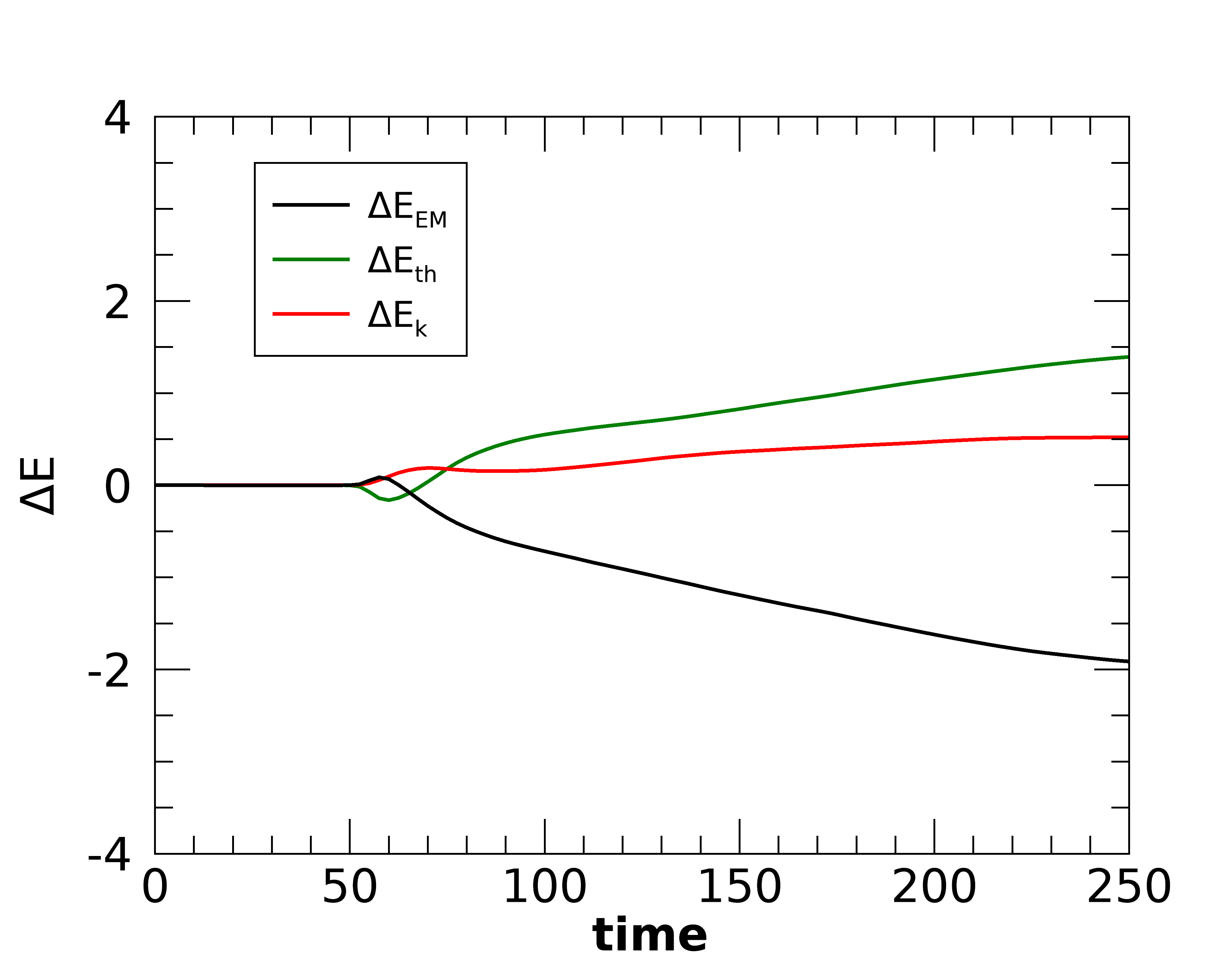}  \end{minipage}
  \hspace{5mm}
  \begin{minipage}{0.47\textwidth}
    \centering    
\includegraphics[width=0.89\columnwidth]{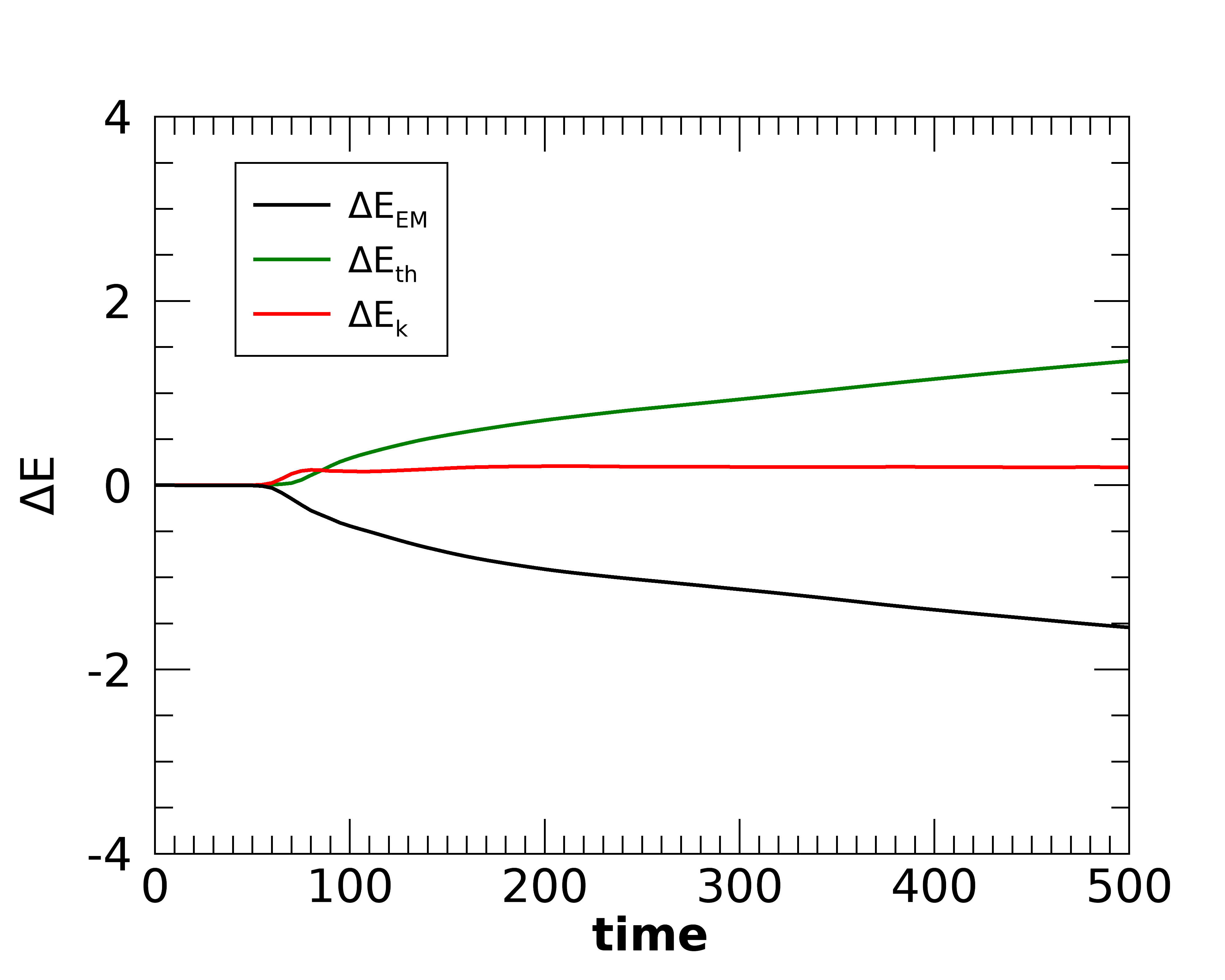}     
  \end{minipage}
    \caption{Evolution of the normalized, kinetic, thermal and electromagnetic energies integrated over the computational domain  with their initial values subtracted, $\Delta E_k$, $\Delta E_{th}$ and  $\Delta E_{em}$ (see text). The top panel refers to the PB case, the middle panel to the CHI06 case while the bottom panel to the FF case.  In the PB case, we also show the results of the simulation PB1 with half the resolution (dashed curve) and the simulation PB2 that makes use of the Taub-Matthews equation of state (dashed-dotted curve)}.
    \label{fig:energies}
\end{figure}

As the evolution proceeds, the tracer acquires values between $0$ and $1$ as a result of the mixing between the column and its environment. 
The value of the tracer in a given cell indicates the percentage of the material that originally was inside the magnetized column. 
In the initial phases, the sharp transition at the jet interface is smoothed out by numerical diffusion and this leads to the initial slow decrease of $M_f$ observed mainly for the CHIO6 and FF cases. 
The decrease is more pronounced for the higher value of the threshold ($f_{th} = 0.5$).  
At later times, the instability evolution  leads to the development of turbulence and hence to a faster turbulent mixing, resulting in the steeper decrease of $M_f$ seen in this figure. 
In the PB case, the evolution is very fast, so that the early exponential growth phase of the instability cannot be captured and already at $t \sim 100$ the initial jet column has been dispersed in the ambient medium. 
The more efficient mixing in the PB case can be related to the growth of short wavelength modes due to PDI (Fig. \ref{fig:linear2}).

The evolution of the instability leads to the energy conversion between electromagnetic, kinetic and thermal energies, in particular, we observe the  formation of dissipative structures in which the magnetic energy is converted mainly into thermal energy. To study this quantitatively, we split the total energy into three parts 
\begin{equation}
    E_{tot} = E_{k} + E_{th} + E_{em}
\end{equation}
where 
\begin{equation*}
E_{k} = \rho\gamma(\gamma -1),~~~~
E_{th} = \gamma^2(w-\rho)-p,~~~~
E_{em} = \dfrac{1}{2}(B^2+E^2)
\end{equation*}
are, respectively, the kinetic, thermal and electromagnetic energy densities. In Fig. \ref{fig:energies} we plot the quantities $\Delta E_k$, $\Delta E_{th}$ and  $\Delta E_{em}$  as a function of time, which  represent, respectively, the integrals of $E_k$, $E_{th}$ and $E_{em}$ over the computational domain, subtracted their corresponding initial values at $t=0$ and normalized by the initial total electromagnetic energy $E_{em,j}$ within the bulk (i.e., for $r\leq a$) of the jet column. The three panels refer to the evolution of these quantities in the PB, CHI06 and FF cases. 

We can observe that the electromagnetic energy decreases and is being converted mainly into thermal energy and,  to a smaller fraction, into kinetic energy. After an  initial readjustment phase, in which the kinetic energy may grow faster than the thermal energy (CHI06 and FF cases) and the thermal energy may even decrease (CHI06 case), the dissipation  reaches a peak and then continues at a smaller rate in the subsequent turbulent phase. This energy conversion process is most efficient in the PB case where, at $t = 130$, we observe the conversion of about $\sim 3.5 E_{em,j}$ of electromagnetic energy into  $70\%$ of thermal and $30\%$ of kinetic energies. Both the efficiency of electromagnetic energy conversion and the percentage that goes into kinetic energy  decrease as we move towards the FF case. 
Specifically, in the CHI06 case, at $t \sim 250$, we have a conversion of $\sim 2 E_{em,j}$ of electromagnetic energy into $75\%$ of thermal and $25 \%$ of kinetic energies and, finally, in the FF case, at $t \sim 500$ we have a conversion of about $\sim 1.5 E_{em,j}$ of electromagnetic energy into $87\%$ of thermal and $13\%$ of kinetic energies. 

 For the PB case we also show  the results for the two additional simulations PB1 and PB2 (Table 1) in Fig. \ref{fig:energies} that have been performed in order to investigate, respectively, the dependence of $\Delta E_k$, $\Delta E_{th}$ and  $\Delta E_{em}$ on the resolution (dashed curve) and on the equation of state (dashed-dotted curve). It is evident that the differences in the evolution of these quantities both for PB1 and PB2 cases with the reference case PB are very small. First of all, this evolution in the PB2  case shows that the choice of the equation of state has a little impact on the results (see also Paper I). As for the dependence on the resolution, different quantities behave differently with resolution, as we already showed in Paper I for the FF case. Comparing the evolution in the PB and lower resolution PB1 cases shown together in the top panel of Fig. \ref{fig:energies}, we conclude that the convergence is reached  for all three quantities plotted in this figure and hence the amount of dissipated energy. By contrast, other quantities, such as the strength of currents within the current sheets keep increasing with resolution, as shown below. However, this is compensated by a reduction of the volume (mostly thickness perpendicular to its surface) of current sheets as the resolution is increased, so that the total amount of dissipated energy (above a certain threshold) does not depend on the resolution, as also shown in Paper I for FF case.

\subsection{Current sheets} 

The instability evolution leads to dissipation of electromagnetic energy,  which is primarily localized  in current sheets. The formation and evolution of the current sheets are demonstrated in Fig. \ref{fig:curr}, showing the $y-z$ section of the current density distribution for all the three PB, CHI06 and FF cases at two different times. 
The panels on the left correspond to the time at which dissipation reaches a peak (see also Fig. \ref{fig:Jcs}), while the panels  on the right  refer to the turbulent quasi-steady state. Note that in the PB case, smaller and stronger current sheets form in the early phases than those in the CHI06 and FF cases due to the small-scale nature of PDI, while at later times, in the turbulent phase, similar disordered smaller-scale current sheet structures are observed in all three cases.

To analyze the properties of the current sheets, we should first identify them. To this end, we follow the same approach already used in Paper I, defining the local steepness parameter
\begin{equation}
    s = \frac{j \delta}{B} \,,
\end{equation}
where $j$ and $B$ are, respectively, the magnitudes of the current density and the magnetic field, while $\delta$ is the cell size. 
$s$ represents a measure of the steepness of magnetic field gradients, the larger is $s$, the smaller is the number of grid points that locally resolve the magnetic field gradients. 
We then identify as cells belonging to a current sheet those cells that are characterized by a value of $s$ larger than a given threshold, $s > s_{\rm th}$. 
Based on the results of Paper I, we choose  $s_{\rm th} = 0.2$ (a different choice of this threshold may change the results only quantitatively but not qualitatively). We also remark that the current density is  computed as $\vec{j} = \nabla \times \vec{B}$, neglecting all the relativistic corrections, since the velocities are mildly relativistic.

We can now examine in a more quantitative way the properties of the current sheets.
In Fig. \ref{fig:Jcs} we show the temporal behavior of $j^2$ integrated over the current sheet regions, specifically
\begin{equation}\label{eq:criteria}
    \int_V \alpha j^2 dV \,,~~~~~
    \alpha = \begin{cases}
			1, & s  \geq s_{th}\\
            0, & \text{otherwise} \,.
		 \end{cases}
\end{equation}
The three panels refer to the three different cases. 
If an explicit resistive term was present in the equations, then the energy density dissipation rate would be given by $\eta j^2$, where $\eta$ is the plasma resistivity (see e.g. Mignone et al. 2019). 
Although our model considers an ideal plasma, we can still take $j^2$ as a proxy of the energy dissipation rate and, therefore, the plots in Fig. \ref{fig:Jcs} can give an estimate of the temporal behaviour of the dissipation rate \citep[see Paper I and][for a more detailed discussion of the use of $j^2$ as a proxy for the dissipation rate in an ideal simulation]{Makwana15}. It is seen in this figure that in all  three cases the dissipation rate reaches a high peak due to strong current sheets, which are of smaller-scale in PB case, but larger-scale in the CHI06 and FF cases (left panels of Fig. \ref{fig:curr}), and then decreases more slowly. 
This peak is much stronger for the PB case and decreases towards the FF case. Eventually the flow settles down into a quasi-steady turbulent state characterized by small-scale irregular current sheets (right panels of Fig. \ref{fig:curr}). The magnitude of $\vec{j}$ in these current sheets in each case and hence the resulting total dissipation rates are smaller than those during the corresponding saturation phases and are nearly constant with time. Note, however, that the relaxation in the PB case takes somewhat longer time and the constant dissipation rate is reached much later.

\begin{figure}
\centering
 \includegraphics[width=\columnwidth]{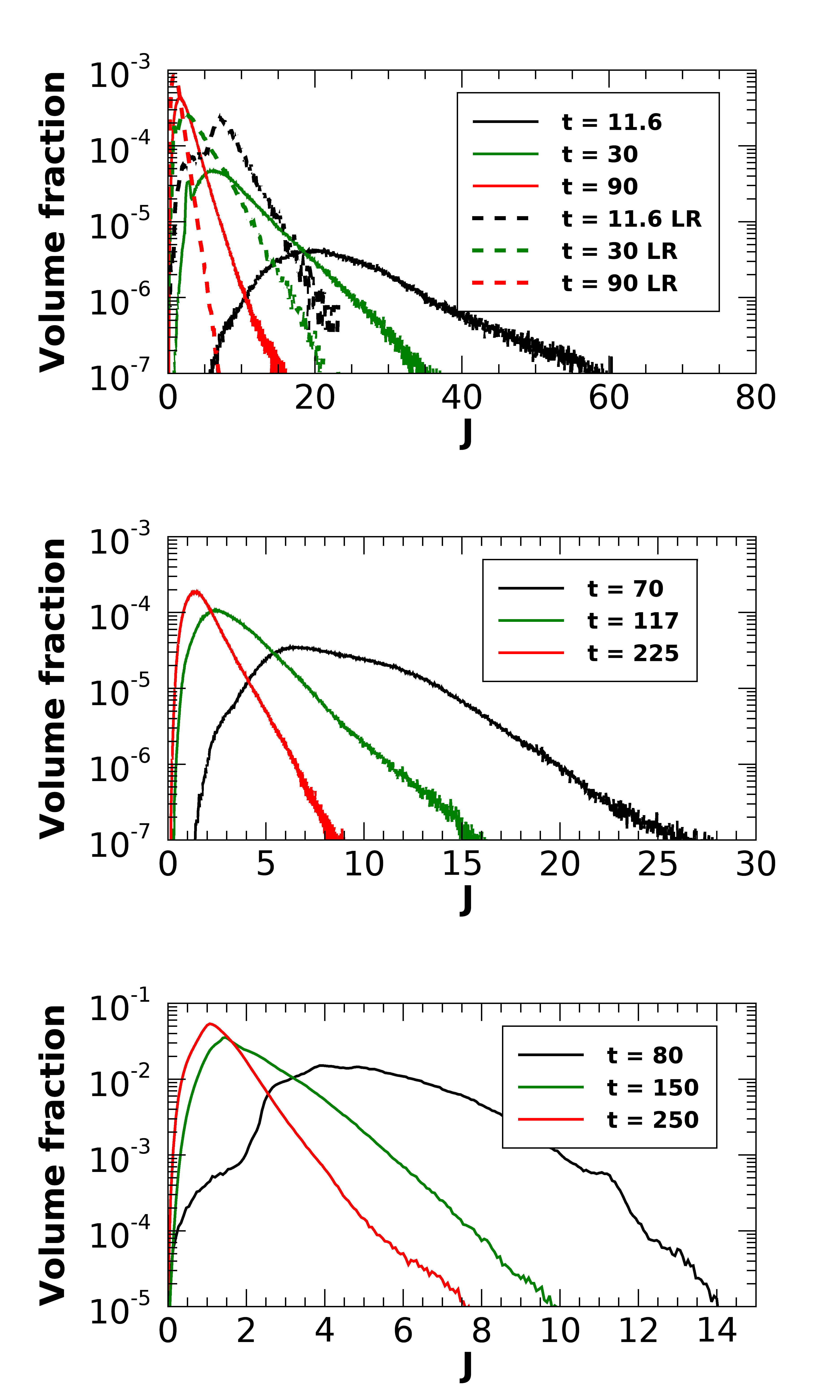}
\caption{Histograms of the volume fraction of current sheets as a function of $j$ for three different times: around the dissipation peak (black), at the end of the peak (green) and in the final quasi-steady saturated turbulent phase (red). The top panel refers to the PB case  (the dashed curves are for the low resolution PB1 case)}, the middle panel to the CHI06 case and the bottom panel to the FF case.
\label{fig:histogram}
\end{figure}
As seen in Fig. \ref{fig:curr}, the decrease in the current sheet strength is related to the widening of the dissipation region as the instability evolves. In Fig. \ref{fig:histogram} we analyze more quantitatively the evolution of the current sheets strength by plotting the histograms of the volume fraction occupied by these current sheets as a function of the current density for the different cases ( the integral of the volume  fraction is normalized to unity for all the curves).
The  three different curves, shown at three different times, correspond to the same main evolutionary stages -- near the peak of the dissipation, end of the peak and finally the fully developed turbulent phase -- represented in Fig. \ref{fig:PB_3D} (the exact time moments are different from those in the latter figure though). The volume fractions, after reaching their maximum, decrease with an exponential tail. During the evolution, both the maximum and the extent of the tail move towards lower values of the current, in agreement with the overall decrease in strength of the current sheets mentioned above.  
Comparing the different cases, we  notice that at early times both the current at the maximum and the extent of the tail are largest in the PB case and, gradually decrease moving to the CHI06 and FF cases. 
In the panel for the PB case, we also show the results for the low resolution PB1 run indicating the effect of lowering (or increasing) resolution on the values of the current. In particular, an increase of resolution leads to an increase in the current density inside the current sheets, which in turn become thinner. An additional point to note is that the highest values of the peaks in the FF case imply that the histograms in this case are more concentrated around the peaks.
At later times, the maxima are found for similar values in each case, but the tails reach the  largest values in the PB case and progressively decrease their extent towards the CHI06 and FF cases. 
This is consistent with the above results reported in Figs. \ref{fig:curr} and \ref{fig:Jcs} that the strength of the current sheets and hence dissipation rate therein decrease moving from the PB to FF cases.

\begin{figure*}
\centering
   \includegraphics[width=0.8\textwidth]{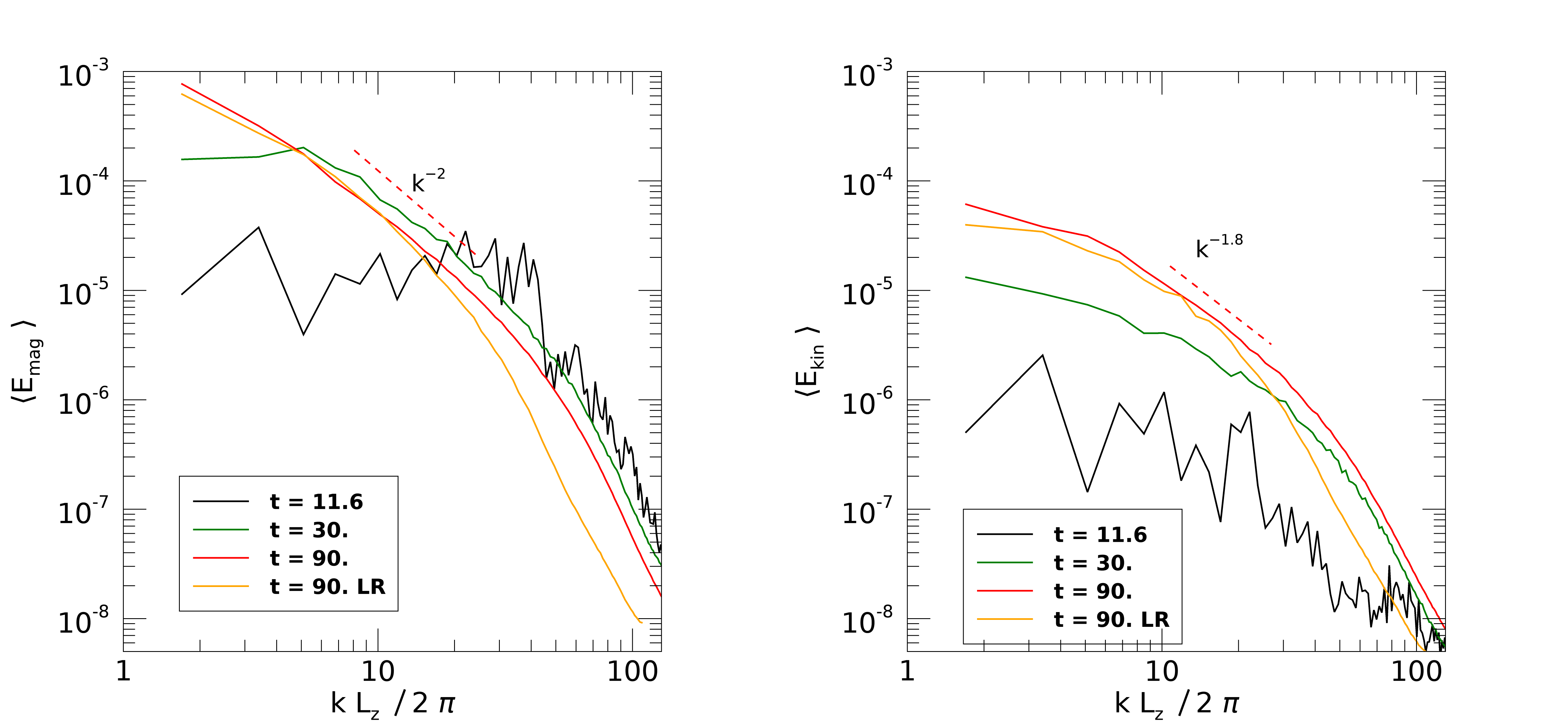}  \\
   \includegraphics[width=0.8\textwidth]{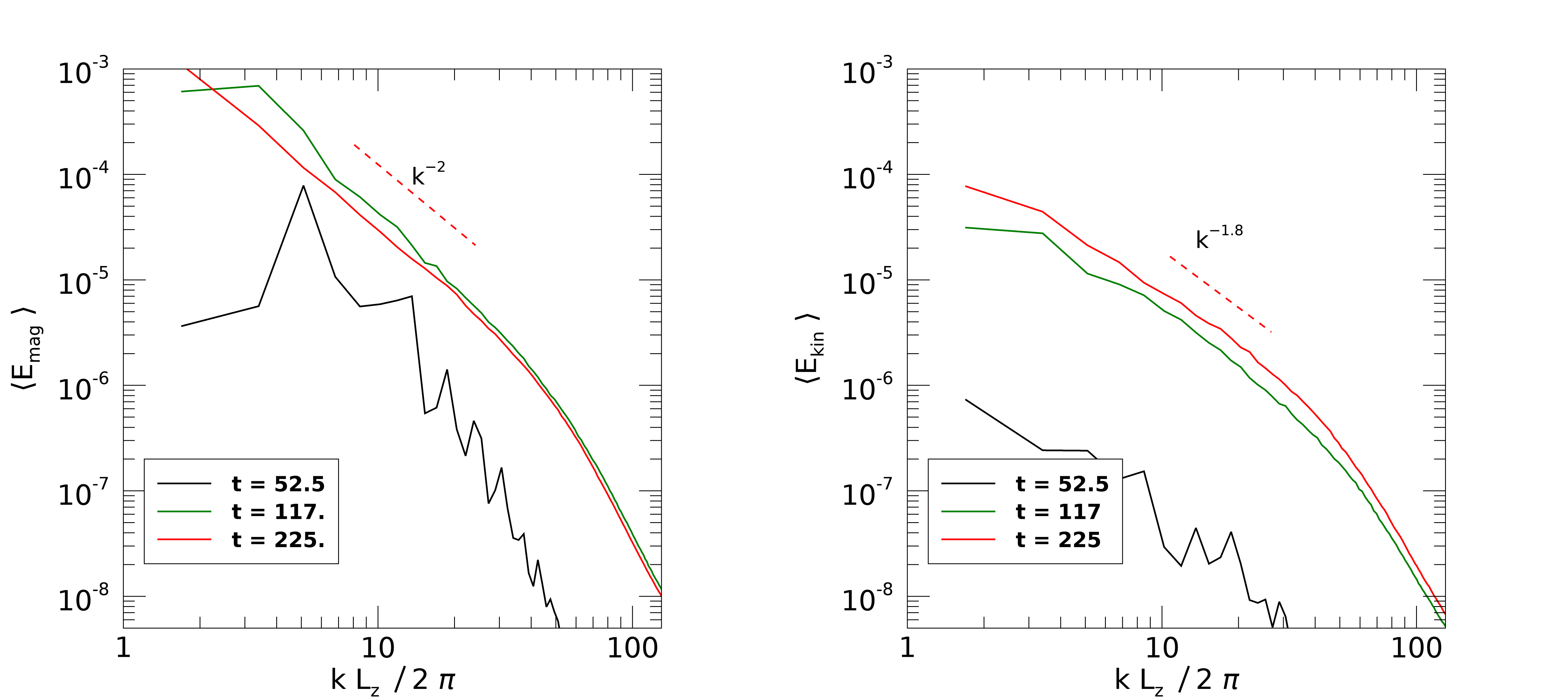} \\
   \includegraphics[width=0.8\textwidth]{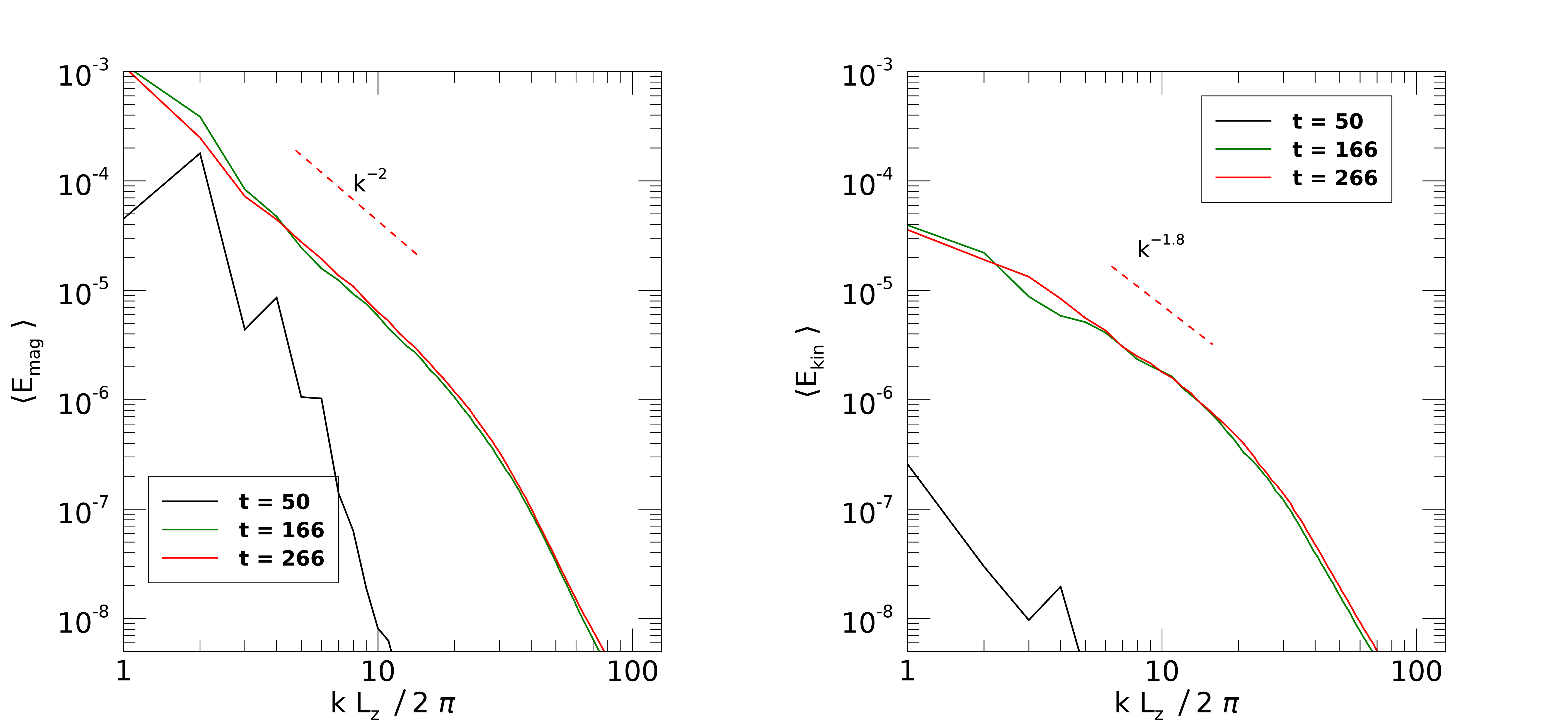}
    \caption{Evolution of the magnetic (left column) and kinetic (right column) energy spectra integrated in the $x-y$ plane and represented as a function of $k$ in the PB (top), CHI06 (middle) and FF (bottom) cases. Black curves correspond to the linear regime, when modes grow exponentially with different growth rates, as a result of PDI, CDI and their mixture in the PB, FF and CHI06 cases, respectively.  Green curves refer to the end of the dissipation peak and the beginning of the nonlinear saturation phase, while the red curves to the final state of the fully developed quasi-steady turbulence. For reference, dashed lines show the power-laws $k^{-2}$ for the magnetic and $k^{-1.8}$ for the kinetic energy spectra For comparison, in the PB case,  we also show the kinetic and magnetic energy spectra in the late turbulent state from the lower resolution run PB1 (orange curves).}\label{fig:spectra}
\end{figure*}


\subsection{Late time evolution -- turbulence}

As we have seen in the above subsections, after the initial burst of the instability, the system settles down into a quasi-steady turbulent state where the dissipation rate is almost constant with time and decreases from the largest value in the PB case to the smallest one in the FF case (Fig. \ref{fig:Jcs}). 
To better understand the nonlinear saturation of CDI and PDI and to characterize the properties of the resulting turbulence, here we analyze its spectral dynamics in the three cases. 
The system is periodic in the longitudinal $z$-direction with a period $L_z$, so we Fourier transform the variables in $z$,
\begin{equation}
   \Bar{g}(x,y,k,t) = \frac{1}{L_z}\int_0^{L_z} g(x,y,z,t) \exp(-i k z) dz  \,,
\end{equation}
where $g = (A, \Vec{B})$, $A\equiv\sqrt{E_k}=\sqrt{\rho \gamma (\gamma - 1)}$ is the square root of the relativistic kinetic energy and $\Vec{B}$ is the magnetic field. Due to the periodicity in $z$ of the computational domain, the wavenumber $k$ is discrete in the simulations, in contrast to the linear analysis, and is determined by the domain size $L_z$, i.e.,  $k = 2 \pi n / L_z$, where the integer $n= 0, \pm 1, \pm 2, ...$. Following Paper I, we define the kinetic $\Bar{E}_{kin}$ and magnetic, $\Bar{E}_{mag}$, energy spectra integrated over the entire $x-y$ plane of the domain
\begin{equation}
\Bar{E}_{kin}(k, t)= \int |\Bar{A}|^2 dx dy,~~~~~~
\Bar{E}_{mag}(k, t) = \frac{1}{2}\int |\Bar{\Vec{B}}|^2 dx dy.
\end{equation}
The temporal evolution of these spectra in the PB, CHI06 and FF cases are shown in Fig. \ref{fig:spectra} at those times that overall correspond to the main stages in Fig. \ref{fig:Jcs} (see also Figs. \ref{fig:PB_3D} and \ref{fig:histogram}). Specifically, the black curves correspond to the exponential growth phase of CDI, PDI and their mixture in the linear regime, somewhat earlier the dissipation peak, respectively, in the FF, PB and CHI06 cases, green curves to the end of the dissipation peak and the beginning of the nonlinear saturation phase and red curves to the final, fully developed turbulent regime. In the first linear regime, the spectra have a ``spiky'' irregular shape, being composed of independently growing unstable modes with different growth rates and wavenumbers. This is in accordance with the dispersion relations for these three cases in Fig. \ref{fig:linear2}, which show that moving from the FF to PB cases, the growth rate of higher $k$ modes gradually increases due to the increasing role of PDI. 
As a result, as seen in Fig. \ref{fig:spectra}, in the linear stage of evolution, the kinetic and magnetic energy spectra, being mostly concentrated at smaller $k$ and the steepest at high $k$ in the FF case, become gradually shallower with increasing $\chi$, reaching comparable magnitudes from smaller to intermediate $1\lesssim kL_z/2\pi \lesssim 40$ in the PB case (larger $kL_z/2\pi \gtrsim 40$ are possibly affected by numerical dissipation and hence cannot grow as fast as lower $k$ modes). In physical space, this trend is manifested in the prevalence of larger-scale structures in the FF and CHI06 cases compared to those in the PB case, as seen from the left panels of Fig. \ref{fig:curr}. The wavenumbers at which the spectra reach the peaks in the CHI06 and FF cases approximately coincide with the respective maxima of the growth rates, that is, with the most unstable modes, in these cases (Fig. \ref{fig:linear2}).

The exponential growth phase ends when the dissipation reaches a maximum (peak) and subsequently the instability enters the nonlinear saturation phase (Fig. \ref{fig:Jcs}). As discussed in the previous subsection, the peak in the dissipation is due to the formation of ordered larger-scale current sheets, where the majority of magnetic dissipation takes place. 
As seen in Fig. \ref{fig:curr}, in the PB case, the current sheets are the smallest, whereas the magnitude of the current density therein is highest. The size of the current sheets and current density decrease when moving to the CHI06 and FF cases. At this time, the role of non-linearity is crucial: it ensures interaction and efficient energy exchange among different (unstable) modes, so that the energy spectra become smooth (green curves in Fig. \ref{fig:spectra}). Larger current sheets start to gradually break up into smaller ones, which correspond to the drop in the total dissipation rate after the peak in Fig. \ref{fig:Jcs}. 
In Fourier space, this process corresponds to direct cascade of energy from small and intermediate unstable $k$ to higher ones due to non-linearity and therefore to the increase (fill-up) of power at these high wavenumbers, hence the spectra appear less steep compared to those in the linear regime. 

After the saturation phase, a quasi-steady turbulent state is established where only small-scale disordered current sheets are left dissipating most of magnetic energy (right panels in Fig. \ref{fig:curr}). The corresponding smooth kinetic and magnetic energy spectra (red curves in Fig. {\ref{fig:spectra}}) are quite close to the spectra during the saturation phase and exhibit a typical signature of a turbulent spectrum -- a power-law dependence at intermediate wavenumbers in the inertial range.  Specifically, the kinetic energy spectra obey the scaling $k^{-1.8}$ and the magnetic energy spectra the scaling $k^{-2}$ at $kL_z/2\pi \lesssim 30$, which, interestingly, are the same in all three cases and are consistent with those obtained only for the FF case in Paper I. This indicates that, regardless of the initial configuration of the jet column, FF, PB or hybrid, in the final turbulent states the spectra at small and intermediate wavenumbers appear to have a similar form and differ only quantitatively, with the FF case having the smallest energies. 
(This is also supported by nearly similar appearance of turbulence structure in physical space in these three cases on the right panels of Fig. \ref{fig:curr}). At higher $kL_z/2\pi\gtrsim 40$, the kinetic and magnetic spectra are steeper due to intensive dissipation in small-scale current sheets at these wavenumbers, scaling as $k^{-4}$ for the kinetic and $k^{-4.5}$ for the magnetic energies. These scalings are the same in these three cases and consistent with those of the turbulent spectra for the FF case from Paper I.

To examine the effect of numerical resolution on the spectral properties, in the top panels of Fig. \ref{fig:spectra} for the PB case, we also show the kinetic and magnetic energy spectra in the late turbulent state from the lower resolution run PB1 (Table 1).\footnote{A detailed resolution study for the FF case is presented in Paper I.} These energy spectra coincide with those of the main higher resolution simulations PB run at small and intermediate $k$ within the inertial range, but decrease steeper at higher $k$, which are affected by resolution (i.e., numerical dissipation). A similar behaviour also holds for the lower resolution spectra in the CHI06 case, so we do not show those spectra here.

\section{Summary and Conclusions}
\label{sec:summary}
In this paper, we have analysed the nonlinear evolution of instabilities in highly magnetized relativistic plasma columns of jets, by means of three-dimensional numerical simulations  for different equilibrium configurations, focusing on the formation of dissipative structures responsible for magnetic energy dissipation. 
 As discussed in Paper I, also in the cases presented in this paper dissipation occurs through the formation of thin current sheets and, as we argued there, its rate appear to depend only on the large-scale characteristics of the flow and not on its small-scale dissipative properties. This is demonstrated by a comparison of the cases with different resolution that, despite their different dissipative properties, show similar dissipation rates. In connection with this it would be important to test this hypothesis  by performing  simulations with explicit resistivity.
Our analysis is relevant for high-energy astrophysical sources since the current sheets that are formed during the instability evolution may be sites of magnetic reconnection, where acceleration of relativistic non-thermal particles   is likely to occur, as shown by PIC simulations \citep[see e.g.][]{Sironi14, Guo14, Werner17, Petropoulou19}.  These simulations show that the acceleration efficiency increases with cold magnetization used here, but it depends also on the plasma $\beta$, i.e., hot magnetization, as shown by \citet{ball18}.
This non-thermal particle population can give rise to the observed high energy emission and thus understanding the dissipation properties becomes of utmost importance for the interpretation of these astrophysical objects \citep[see e.g.][]{Zhang16, Zhang18, Bodo21}. 

The considered equilibrium structures are characterized by different balances between three main radial forces acting on the jet column: the tension due to the azimuthal component of the magnetic field, the magnetic pressure gradient, related to the longitudinal component of the field, and the thermal pressure gradient. 
We compared the results obtained in Paper I for the force-free case with configurations in which the thermal pressure gradient plays a role in the equilibrium force balance. We have different kind of instabilities in the different cases. Specifically, 
 in the FF case, the instability driver is  the current parallel to the field and we have  current-driven instabilities (CDI), while, in the PB case, the instability driver is the current perpendicular to the field and we have pressure-driven instabilities (PDI).
As a complementary results to the numerical simulations, we performed the linear stability analysis of the above equilibria and characterized the properties of CDI, PDI and their mixture in the hybrid case. Our analysis showed that the growth rate of PDI increases with the wavenumber until it reaches a plateau for sufficiently large $k$, both for the axisymmetric $m=0$ and non-axisymmetric $m=1$ kink modes, for which the growth rates are comparable. The decrease in the thermal pressure gradient in the force balance, results in the decrease in the growth rate of the instability and the appearance of a cutoff at high wavenumbers.  In this case, the axisymmetric $m=0$ mode gradually disappears and becomes completely stable in the FF equilibrium, which is thus subject only to the non-axisymmetric $m=1$ kink mode of CDI.

Our results from the nonlinear simulations are in general agreement with those of \citet{Oneill12, Striani16, Ortuno22}.
In the nonlinear evolution of both CDI, PDI and their combination in the hybrid case proceeds in general into two phases. 
At first, the jet  undergoes helicoidal deformation, with the formation  of strong current sheets, where  substantial dissipation of magnetic energy occurs. 
The peak of dissipation is then followed by a quasi-steady turbulent state, where energy dissipation still takes place, but at a much lower rate.  Among the three equilibria, the evolution in the PB case is the fastest and most energetic, as also indicated by the linear analysis. In this case,   the current sheets, resulting from the instability evolution  are characterized by the strongest current density and the  shortest length scale and yield the highest dissipation rate.  PIC simulations performed by \citet{Alves18}, \citet{Davelaar20} and \citet{Ortuno22} have shown that the Z-pinch configuration (our PB equilibrium) is most effective in the particle acceleration process and this is consistent with our finding of a stronger dissipation for this case.

On the contrary, in the FF case, we observe at the beginning of the CDI evolution a dominant unstable mode of larger wavelength, in agreement with the linear analysis. 
As a result, the current sheets that develop at this stage are the weakest and hence the dissipation rate the smallest. 
The hybrid case (CHI06) lies between the two extreme  FF and PB cases, in terms of the time-scale of evolution, the initial dominant wavelength of the instability and the amount of dissipation. 
 All the cases share a similar subsequent evolutionary path -- the development of quasi-steady turbulence with a characteristic power-law behavior for the spectra of the kinetic and magnetic energies. 
In this turbulent state, disordered small-scale, weaker current sheets form via turbulent cascade, or fragmentation of the larger-scale ones present at the earlier stage of the nonlinear development of the instability and are mainly responsible for energy dissipation, but at a much slower rate.

These results are relevant for the interpretation of the phenomenology of high-energy astrophysical sources. 
As already  pointed out, magnetic reconnection is thought to be an important mechanism for the acceleration of the relativistic particles that are at the origin of the observed non-thermal emission in those sources. 
The evolution of both current and pressure driven instabilities  provides a natural stage for the formation of current sheets that may be sites of magnetic reconnection. 
In addition, we may have particle acceleration also in the steady turbulent phase. 
The different  properties of the current sheets in the three cases may have consequences in terms of the emission and polarization characteristics of the observed emission.
The strength of the current sheets may be directly connected to the intensity of the observed radiation, while the dominant wavelength may have consequences for the polarization signatures. 
In order to obtain detailed predictions to be compared with the observational data, one needs to introduce, in the RMHD simulations, sub-grid models for treating particle acceleration and their emission, since these processes occur at a much smaller scale that cannot be reached by RMHD simulations \citep[see e.g.][]{vaidya18}.  
As a first step one can use rather crude sub-grid models like in  \citet{Zhang16}  and \citet{Bodo21}. 
We plan however to introduce, in the next stage of investigation more refined models \citep{Nurisso23}. 
Another limitation of  the present work is the absence of the jet velocity and we plan to overcome also this limitation in our subsequent investigations.

\section*{Acknowledgments}
This project has received funding from the European Research Council (ERC) under the European Union’s Horizon 2020 research and innovation program (Grant Agreement No. 787544) as well as from the European High Performance Computing Joint Undertaking (JU) and Belgium, Czech Republic, France, Germany, Greece, Italy, Norway, and Spain under grant agreement No 101093441 (SPACE). We also acknowledge support from the fondazione ICSC, Spoke 3 Astrophysics and Cosmos Observations and National Recovery and Resilience Plan (Piano Nazionale di Ripresa e Resilienza, PNRR) Project ID CN\_00000013 \quotes{Italian Research Center on High-Performance Computing, Big Data and Quantum Computing}  funded by MUR Missione 4 Componente 2 Investimento 1.4: Potenziamento strutture di ricerca e creazione di \quotes{campioni nazionali di $R\&S$ (M4C2-19 )} - Next Generation EU (NGEU).  This work was also partly supported by PRIN MUR 2022 (grant n. 2022C9TNNX)  and by the INAF Theory Grant "Multi scale simulations of relativistic jets".  We acknowledge the CINECA ISCRA program and  the INAF-CINECA {\it Accordo Quadro MoU per lo svolgimento di attività congiunta di ricerca Nuove frontiere in Astrofisica: HPC e Data Exploration di nuova generazione}, for the availability of computing resources.

\section{Data availability}
The data underlying this article will be shared on reasonable request
to the corresponding author

\bibliographystyle{mn2e}
\bibliography{main.bib}

\appendix
\section{Coefficients of the linear system}

In the linear analysis based on equations (\ref{eq:linear1}) and (\ref{eq:linear2}), we normalize displacement $\xi_{1r}$ by the magnetization radius $a$ of the jet (which is the unit of length in the paper). For convenience, in the linear equations below we use the equilibrium profiles of the magnetic field components $B_{\varphi}$ and $B_{z}$, divided by $B_0$ and pressure $p$ divided by $B_0^2$,
\[
\mathfrak{B}_{\varphi}^{2}(r)=\frac{1}{r^{2}}\left[1-\exp\left(-r^{4}\right)\right],~~~~~\mathfrak{B}_{z}^{2}(r)=P_{c}^2-\left(1-\chi\right)\sqrt{\pi}\mathrm{\,erf}\left(r^{2}\right),
\]
\[
\mathfrak{B}^{2}=\mathfrak{B}_{\varphi}^{2}+\mathfrak{B}_{z}^{2},~~~~~~\mathcal{P}(r)=\frac{p_a}{B_0^2}+\chi\frac{\sqrt{\pi}}{2}\left[1-\mathrm{erf}\left(r^{2}\right)\right].
\]
In terms of these normalized equilibrium profiles, the coefficients in the system of linear equations (\ref{eq:linear1}) and (\ref{eq:linear2}) can be derived with lengthy algebra and have the form,
\begin{equation}\label{eq:A11}
A_{11}=-\left(\frac{1}{K_{\parallel}}\frac{dK_{\parallel}}{dr}+\frac{1}{\mathfrak{B}}\frac{d\mathfrak{B}}{dr}+\frac{1}{r}\right)-\frac{m}{rK_{\parallel}}\frac{1}{\mathfrak{B}}\mathrm{\mathcal{C}_{1}}-\frac{k}{K_{\parallel}}\frac{1}{\mathfrak{B}}\mathrm{\mathcal{C}_{2},}
\end{equation}
\begin{equation}\label{eq:A12}
A_{12}=-\frac{m}{rK_{\parallel}}\frac{1}{\mathfrak{B}}\mathrm{\mathcal{C}_{3}}-\frac{k}{K_{\parallel}}\frac{1}{\mathfrak{B}}\mathrm{\mathcal{C}_{4}},
\end{equation}
\begin{equation}\label{eq:A21}
A_{21}=\mathfrak{B}^{2}\left(\frac{\omega^{2}}{V_{A}^{2}}-K_{\parallel}^{2}\right)-2\frac{\mathfrak{B}_{\varphi}}{r}\mathrm{\mathcal{C}_{1}},
\end{equation}
\begin{equation}\label{eq:A22}
A_{22}=-2\frac{\mathfrak{B}_{\varphi}}{r}\mathrm{\mathcal{C}_{3}},
\end{equation}
where we used the following definitions

\begin{multline*}
\mathcal{C}_1 = \frac{1}{\mathfrak{B}^2 \mathfrak{D}_1  \mathfrak{D}_2}\left[\mathfrak{D}_2\mathfrak{B}_{\varphi}\mathcal{P}\left(\mathfrak{D}_{1}\frac{1}{\mathcal{P}}\frac{d\mathcal{P}}{dr}+\frac{2\Gamma}{\mathfrak{B}^2}\frac{\mathfrak{B}_{\varphi}^2}{r}\right)+\right.\\\left.\mathfrak{B}_{z}\mathfrak{D}_{1}\left(2\omega^{2}\left(1+\mathcal{S}\right)\frac{\mathfrak{B}_{z}\mathfrak{B}_{\varphi}}{r}-\mathfrak{D}_{2}J_{\parallel}\right)\right],
\end{multline*}

\begin{multline*}
\mathcal{C}_2 = \frac{1}{\mathfrak{B}^2 \mathfrak{D}_1  \mathfrak{D}_2}\left[\mathfrak{D}_2\mathfrak{B}_{z}\mathcal{P}\left(\mathfrak{D}_{1}\frac{1}{\mathcal{P}}\frac{d\mathcal{P}}{dr}+\frac{2\Gamma}{\mathfrak{B}^2}\frac{\mathfrak{B}_{\varphi}^2}{r}\right)-\right. \\\left. \mathfrak{B}_{\varphi}\mathfrak{D}_{1}\left(2\omega^{2}\left(1+\mathcal{S}\right)\frac{\mathfrak{B}_{z}\mathfrak{B}_{\varphi}}{r}-\mathfrak{D}_{2}J_{\parallel}\right)\right],
\end{multline*}

\begin{equation*}
\mathcal{C}_3 = \frac{1}{\mathfrak{B}^2 \mathfrak{D}_1  \mathfrak{D}_2}\left[\mathfrak{D}_{2}\left(1-\frac{C_{s}^{2}K_{\parallel}^2}{\omega^{2}}\right)\mathfrak{B}_{\varphi}+\mathcal{S}K_{\parallel}K_{\perp}\mathfrak{B}_{z}\mathfrak{D}_{1}  \right],
\end{equation*}

\begin{equation*}
\mathcal{C}_4 = \frac{1}{\mathfrak{B}^2\mathfrak{D}_1  \mathfrak{D}_2} \left[\mathfrak{D}_2\left(1-\frac{C_{s}^{2}K_{\parallel}^2}{\omega^{2}}\right)\mathfrak{B}_{z}-\mathcal{S}K_{\parallel}K_{\perp}\mathfrak{B}_{\varphi}\mathfrak{D}_{1}\right],
\end{equation*}
\[
\mathfrak{D}_{1}=1+\frac{\Gamma\mathcal{P}}{\mathfrak{B}^{2}}-\frac{C_s^2K_{\parallel}^{2}}{\omega^{2}},\qquad\mathfrak{D}_{2}=\mathcal{S}\left(\frac{\omega^{2}}{V_{A}^{2}}-K_{\parallel}^{2}\right),
\]
\[
K_{\parallel}=\frac{k\mathfrak{B}_{z}+m\mathfrak{B}_{\varphi}/r}{\mathfrak{B}},\qquad K_{\perp}=\frac{k\mathfrak{B}_{\varphi}-m\mathfrak{B}_{z}/r}{\mathfrak{B}},
\]
with $\mathcal{S}$, characterizing the magnetization as a function of radius,
\[
\mathcal{S}=\frac{B_{0}^{2}\mathfrak{B}^{2}}{\rho_{0}+\Gamma B_{0}^{2}\mathcal{P}/\left(\Gamma-1\right)},
\]
the sound speed squared
\[
C_{s}^{2}=\frac{\Gamma B_{0}^{2}\mathcal{P}}{\rho_{0}+\Gamma B_{0}^{2}\mathcal{P}/\left(\Gamma-1\right)},
\]
and the Alfv\'en speed squared
\[
V_{A}^{2}=\frac{B_{0}^{2}\mathfrak{B}^{2}}{B_{0}^{2}\mathfrak{B}^{2}+\rho_{0}+\Gamma B_{0}^{2}\mathcal{P}/\left(\Gamma-1\right)}=\frac{\mathcal{S}}{1+\mathcal{S}},
\]
where $\Gamma$ is the adiabatic index.

The projection of the current parallel to the background magnetic field is
\[
J_{\parallel}=\mathfrak{B}_{\varphi}\mathfrak{J}_{\varphi}+\mathfrak{B}_z\mathfrak{J}_z=\frac{\mathfrak{B}_{z}}{r}\frac{\partial}{\partial r}\left(r\mathfrak{B}_{\varphi}\right)-\mathfrak{B}_{\varphi}\frac{\partial\mathfrak{B}_{z}}{\partial r}.
\]

\section{Asymptotic solution at small radii}

\subsection{The $m\neq 0$ case}
In this section, we derive boundary conditions at small radii $r\rightarrow 0$ first for non-axisymmetric, $m\neq 0$, modes  by expanding the coefficients $\mathrm{A}_{11},~\mathrm{A}_{12},~\mathrm{A}_{21},~ \mathrm{A}_{22}$ from Appendix A around the origin $r=0$. Before that, we expand the equilibrium quantities around $r=0$, keeping only the terms up to order $r^2$,
\[
\mathfrak{B}_{\varphi}=-r <0,~~~~\mathfrak{B}_z=P_c-\frac{1-\chi}{P_c}r^2>0~~({\rm pitch~is~positive~ }P_c>0),
\]
\[
\mathcal{P}=\frac{p_a}{B_0^2}+\chi\frac{\sqrt{\pi}}{2}-\chi r^2,~~~~
K_{\parallel}=\frac{k\mathfrak{B}_z-m}{\mathfrak{B}_z},~~~~K_{\perp}=-\frac{m}{r},~~~~J_{\parallel}=-2\mathfrak{B}_z.
\]
With these expansions we derive $\mathcal{C}_1$, $\mathcal{C}_2$, $\mathcal{C}_3$, $\mathcal{C}_4$ to leading order in small $r$,
\[
\mathcal{C}_1=-\left.\frac{2V_A^2K_{\parallel}^2}{\omega^2-V_A^2K_{\parallel}^2}\right\vert_{r=0}
\]
\[
\mathcal{C}_2=\frac{2}{\mathfrak{B}_z\mathfrak{D}_1
\mathfrak{D}_2}\left[\mathfrak{D}_2\left(\frac{\Gamma\mathcal{P}}{\mathfrak{B}_z^2}-\chi\mathfrak{D}_1\right)-\mathcal{S}\mathfrak{D}_1K_{\parallel}^2\right]_{r=0}\cdot r
\]
\[
\mathcal{C}_3=-\left.\frac{mK_{\parallel}V_A^2}{\mathfrak{B}_z(\omega^2-K_{\parallel}^2V_A^2)}\right\vert_{r=0}\cdot\frac{1}{r},
\]
\[
\mathcal{C}_4 = \frac{1}{\mathfrak{B}_z^2\mathfrak{D}_1\mathfrak{D}_2}\left[\mathfrak{D}_2\mathfrak{B}_{z}\left(1-\frac{C_{s}^{2}K_{\parallel}^2}{\omega^{2}}\right)-m\mathcal{S}K_{\parallel}\mathcal{D}_1\right]_{r=0}.
\]
Using these expressions, we can compute now the coefficients $\mathrm{A}_{11},~\mathrm{A}_{12},~\mathrm{A}_{21},~\mathrm{A}_{22}$ from expressions (\ref{eq:A11})-(\ref{eq:A22}) also to leading order in $r$,
\[
\mathrm{A}_{11}=\left[\frac{2mV_A^2K_{\parallel}}{\mathfrak{B}_z(\omega^2-V_A^2K_{\parallel}^2)}-1 \right]_{r=0}\cdot\frac{1}{r}=\frac{a_{11}}{r},
\]
\[
\mathrm{A}_{12}=\left.\frac{m^2V_A^2}{\mathfrak{B}_z^2(\omega^2-V_A^2K_{\parallel}^2)}\right\vert_{r=0}\cdot \frac{1}{r^2}=\frac{a_{12}}{r^2},
\]
\[
\mathrm{A}_{21}=\left.\mathfrak{B}_z^2\left(\frac{\omega^2}{V_A^2}-K_{\parallel}^2\right)-\frac{4V_A^2K_{\parallel}^2}{\omega^2-V_A^2K_{\parallel}^2}\right\vert_{r=0}=a_{21},
\]
\[
\mathrm{A}_{22}=-\left.\frac{2mK_{\parallel}V_A^2}{\mathfrak{B}_z(\omega^2-V_A^2K_{\parallel}^2)}\right\vert_{r=0}\cdot\frac{1}{r}=\frac{a_{22}}{r},
\]
where we have introduced the auxiliary coefficients $a_{11},~a_{12},~ a_{21},~a_{22}$, which are all calculated at $r=0$ and are finite, in order to better separate out the dependence of the original coefficients $\mathrm{A}_{11},~\mathrm{A}_{12},~\mathrm{A}_{21},~\mathrm{A}_{22}$ on small $r$. These new coefficients will be convenient in the following derivations. Using these notations equations (\ref{eq:linear1}) and (\ref{eq:linear2}) become,
\begin{equation}\label{eq:linear1_reduced}
    \frac{d \xi_{1r}}{dr} = \frac{a_{11}}{r}\xi_{1r} +\frac{a_{12}}{r^2} \Pi_1
\end{equation}
\begin{equation}\label{eq:linear2_reduced}
    \frac{d \Pi_1}{dr} = a_{21} \xi_{1r}+\frac{a_{22}}{r}\Pi_1.
\end{equation}
We look for solutions to this system in the form 
$$\xi_{1r}=Ar^{\alpha}, ~~~~~\Pi_1=Br^{\alpha+1},$$ 
where $A$ and $B$ are the arbitrary coefficients and $\alpha$ is some exponent to be determined. Substituting this solution into equations (\ref{eq:linear1_reduced}) and (\ref{eq:linear2_reduced}), yields the quadratic equation for $\alpha$
\[
\alpha^2-\alpha(a_{11}+a_{22}-1)+a_{11}a_{22}-a_{12}a_{21}-a_{11}=0,
\]
which after some algebra simplifies to
\[
\alpha^2+2\alpha+1-m^2=0
\]
with the roots
\[
\alpha=\pm |m|-1.
\]
A solution must be regular (finite) as $r \rightarrow 0$, therefore we choose only the root with a plus sign, $\alpha = |m| - 1,~ (|m| \geq 1)$. These roots for $\alpha$ and hence the behaviour of solution near $r=0$ are similar to those derived in Appendix C of \citet{Bodo13} in the case of a cold jet. As a result, for the ratio of $A$ and $B$ coefficients in $\xi_{1r}$ and $\Pi_1$ near $r=0$, we get
\[
\frac{B}{A}=\frac{a_{21}}{|m|-a_{22}}=\frac{\mathfrak{B}_z}{m}\left[{\rm sign}(m)\mathfrak{B}_z\left(\frac{\omega^2}{V_A^2}-K_{\parallel}^2\right)-2K_{\parallel}\right]_{r=0},
\]
with $\alpha=|m|-1$, which determines the solution at small radii. This solution is then used in the shooting method to integrate $\xi_{1r}$ and $\Pi_1$ farther from $r=0$ along radius and connect to the solution coming from large radii.

\subsection{The $m=0$ case}

For axisymmetric $m=0$ modes, the leading order terms in the expansion of $K_{\perp}$ and $\mathcal{C}_3$ in $r$ that are proportional to $m$ vanish and an expansion to the next order is necessary, while in $K_{\parallel}$ and $\mathcal{C}_4$, which also depend on $m$, we simply put $m=0$,
\[
K_{\parallel}=k,~~~~K_{\perp}=k\frac{\mathfrak{B}_{\varphi}}{\mathfrak{B}_z},~~~~\mathcal{C}_4=\left.\frac{1}{\mathfrak{B}_z\mathfrak{D}_1}\left(1-\frac{C_s^2K_{\parallel}^2}{\omega^2}\right)\right\vert_{r=0}
\]
\[
\mathcal{C}_3=-\frac{\mathcal{S}}{\mathfrak{B}_z^2\mathfrak{D}_1\mathfrak{D}_2}\left[\frac{\omega^2}{V_A^2}+K_{\parallel}^2\left(\frac{\Gamma\mathcal{P}}{\mathfrak{B}_z^2}-\frac{C_s^2}{V_A^2}\right)\right]_{r=0}
\cdot r.
\]
Taking this into account, for the coefficients $\mathrm{A}_{11}$, $\mathrm{A}_{12}$, $\mathrm{A}_{21}$, $\mathrm{A}_{22}$ to leading order in $r$, we get
\[
\mathrm{A}_{11}=-\frac{1}{r},~~~~\mathrm{A}_{12}=-\left.\frac{1}{\mathfrak{B}_z^2\mathfrak{D}_1}\left(1-\frac{C_s^2K_{\parallel}^2}{\omega^2}\right)\right\vert_{r=0}=b_{12},
\]
\[
\mathrm{A}_{21}= \left.\mathfrak{B}_z^2\left(\frac{\omega^2}{V_A^2}-K_{\parallel}^2\right)-\frac{4V_A^2K_{\parallel}^2}{\omega^2-V_A^2K_{\parallel}^2}\right\vert_{r=0}=b_{21},
\]
\[
\mathrm{A}_{22}=-\frac{2\mathcal{S}}{\mathfrak{B}_z^2\mathfrak{D}_1\mathfrak{D}_2}\left[\frac{\omega^2}{V_A^2}+K_{\parallel}^2\left(\frac{\Gamma\mathcal{P}}{\mathfrak{B}_z^2}-\frac{C_s^2}{V_A^2}\right)\right]_{r=0}
\cdot r=b_{22}\cdot r.
\]
In these expressions, we have again introduced the auxiliary constants $b_{12},~b_{21},~b_{22}$ calculated at $r=0$ for the convenience of the following derivations. With these notations we write equations (\ref{eq:linear1}) and (\ref{eq:linear2}) as
\begin{equation}\label{eq:linear1_m0}
    \frac{d \xi_{1r}}{dr} = -\frac{\xi_{1r}}{r} + b_{12} \Pi_1
\end{equation}
\begin{equation}\label{eq:linear2_m0}
    \frac{d \Pi_1}{dr} = b_{21} \, \xi_{1r} + b_{22}r \Pi_1.
\end{equation}
Expressing $\xi_{1r}$ from equation (\ref{eq:linear2_m0}), substituting into equation (\ref{eq:linear1_m0}) and keeping only the leading order terms in $r$, we get a single second order ordinary differential equation for $\Pi_1$ only,
\[
\frac{d^2\Pi_1}{dr^2}+\frac{1}{r}\frac{d\Pi_1}{dr}-(2b_{22}+b_{12}b_{21})\Pi_1=0.
\]
This equation has the form of the modified Bessel equation 
\[
\frac{d^2\Pi_1}{dr^2}+\frac{1}{r}\frac{d\Pi_1}{dr}-\mu^2\Pi_1=0, 
\]
where $\mu^2\equiv 2b_{22}+b_{12}b_{21}$. Its solution which is finite and well-behaved at $r\rightarrow 0$ is given by the modified Bessel function of the first kind $I_0(\mu r)$,
\[
\Pi_1=BI_0(\mu r)=B\left(1+\frac{\mu^2r^2}{4}+\frac{\mu^4r^4}{64}+...\right),
\]
where $B$ is an arbitrary constant. Next, expressing the displacement from equation (\ref{eq:linear2_m0}), we get
\[
\xi_{1r}=\frac{1}{b_{21}}\left(\frac{d\Pi_1}{dr}-b_{22}r\Pi_1 \right)=B\left(\frac{b_{12}}{2}r+\frac{b_{12}^2b_{21}^2-4b_{22}^2}{16b_{21}}r^3+...\right).
\]
These expressions for $\Pi_1$ and $\xi_{1r}$ are used for setting boundary conditions for axisymmetric $m=0$ modes at small $r$.

\section{Asymptotic solution at large radii}

Here we derive the boundary conditions at large radii, $r\rightarrow \infty$. For this we first calculate asymptotic behavior of each term $\mathrm{A}_{11}, \mathrm{A}_{12}, \mathrm{A}_{21}, \mathrm{A}_{22}$ given in Appendix A at large radii. For the equilibrium quantities at $r\rightarrow \infty$, we have 
\[
K_{\parallel}=k,~~~~~K_{\perp}= -m/r,~~~~~ \mathfrak{B}_{\varphi}=0,
\]
\[
\mathfrak{B}=\mathfrak{B}_z= P_c^2-(1-\chi)\sqrt{\pi},~~~~~\mathcal{P}= p_a/B_0^2,~~~~~J_{\parallel}=0,
\]
which except for $K_{\perp}$ do not vary with $r$. Using these expressions, we find  $\mathcal{C}_1$, $\mathcal{C}_2$, $\mathcal{C}_3$, $\mathcal{C}_4$ up to terms of order $1/r$,
\[
\mathcal{C}_1=0,~~~\mathcal{C}_2=0,~~~\mathcal{C}_3= -\frac{\mathcal{S}mk}{\mathfrak{B}_z\mathfrak{D}_2}\cdot \frac{1}{r},~~~
\mathcal{C}_4 = \frac{1}{\mathfrak{B}_z\mathfrak{D}_1}\left(1-\frac{C_s^2k^2}{\omega^2}\right) 
\]
and therefore for the coefficients $\mathrm{A}_{11}, \mathrm{A}_{12}, \mathrm{A}_{21}, \mathrm{A}_{22}$ we get
\[
\mathrm{A}_{11}= -\frac{1}{r},~~~~~~\mathrm{A}_{12}= \frac{\mathcal{S}}{\mathfrak{B}_z^2\mathfrak{D}_2}\frac{m^2}{r^2}-\frac{1}{\mathfrak{B}_z^2\mathfrak{D}_1}\left(1-\frac{C_s^2k^2}{\omega^2}\right)
\]
\[
\mathrm{A}_{21}=\mathfrak{B}_z^2\left(\frac{\omega^2}{V_A^2}-k^2 \right),~~~~\mathrm{A}_{22}= 0, 
\]
where all the quantities in these expressions are taken at $r\rightarrow \infty$. Substituting them into equations (\ref{eq:linear1}) and (\ref{eq:linear2}), we obtain
\begin{equation}\label{eq:linear1_infty}
    \frac{d \xi_{1r}}{dr} = -\frac{\xi_{1r}}{r} + \mathrm{A}_{12} \Pi_1
\end{equation}
\begin{equation}\label{eq:linear2_infty}
    \frac{d \Pi_1}{dr} = \mathrm{A}_{21} \, \xi_{1r}.
\end{equation}
Expressing $\xi_{1r}$ from equation (\ref{eq:linear2_infty}) and substituting into equation (\ref{eq:linear1_infty}), we arrive at a single second order differential equation only for $\Pi_1$,
\begin{equation}\label{eq:equationpi}
\frac{d^2\Pi_1}{dr^2}+\frac{1}{r}\frac{d\Pi_1}{dr}-A_{12}A_{21}\Pi_1=0,
\end{equation}
or after substituting explicit expressions for $\mathrm{A}_{12}$ and $\mathrm{A}_{21}$,
\begin{equation*}
\frac{d^2\Pi_1}{dr^2}+\frac{1}{r}\frac{d\Pi_1}{dr}+\left[\frac{1}{\mathfrak{D}_1}\left(\frac{\omega^2}{V_A^2}-k^2\right)\left(1-\frac{C_s^2k^2}{\omega^2}\right)-\frac{m^2}{r^2}\right]\Pi_1=0.
\end{equation*}
This equation has the form of Bessel equation
\begin{equation}\label{eq:bessel_pi}
\frac{d^2\Pi_1}{dr^2}+\frac{1}{r}\frac{d\Pi_1}{dr}+\left(\beta^2-\frac{m^2}{r^2}\right)\Pi_1=0,
\end{equation}
where the complex constant $\beta$ is given by
\[
\beta^2=\frac{1}{\mathfrak{D}_1}\left(\frac{\omega^2}{V_A^2}-k^2\right)\left(1-\frac{C_s^2k^2}{\omega^2}\right).
\]
Solution of equation (\ref{eq:bessel_pi}) corresponding to propagating waves that vanish at infinity is given by Hankel function of the first kind, $\Pi_1=H_{m}^{(1)}(\beta r)$, with the leading term of asymptotic expansion at $r\rightarrow \infty$
\begin{equation}\label{eq:solution_pi}
\Pi_1=H_{m}^{(1)}(\beta r)\simeq \sqrt{\frac{2}{\pi \mu r}}{\rm exp}\left[i\left(\beta r-\frac{m\pi}{2}-\frac{\pi}{4}\right)\right].
\end{equation}
Note that this solution is general, valid both for non-axisymmetric, $m\neq 0$, and axisymmetric, $m=0$, modes. 

The complex parameter $\beta$ can have both signs,
\[
\beta=\pm\sqrt{\frac{1}{\mathfrak{D}_1}\left(\frac{\omega^2}{V_A^2}-k^2\right)\left(1-\frac{C_s^2k^2}{\omega^2}\right)}.
\]
The wave perturbations should decay at large radii, therefore $\beta$ with a positive imaginary part, ${\rm Im}(\beta) > 0$, should be chosen. These waves originate in the jet and propagate radially outwards in the form of outgoing waves at large radii. This means that the real parts of $\beta$ and $\omega$ should have opposite signs, ${\rm Re}(\omega){\rm Re}(\beta) < 0$ (Sommerfeld condition), resulting in the phase velocity directed radially outwards.

The asymptotic behaviour of $\xi_{1r}$ can be readily obtained from equation (\ref{eq:linear2_infty}),
\begin{equation}\label{eq:solution_xi}
\xi_{1r}=\frac{1}{\mathrm{A}_{21}}\frac{d\Pi_1}{dr}\simeq \frac{V_A^2}{\mathfrak{B}_z^2(\omega^2-V_A^2k^2)}\left(i\beta-\frac{1}{2r}\right)\Pi_1 
\end{equation}
The asymptotic solutions (\ref{eq:solution_pi}) and (\ref{eq:solution_xi}) are used as an initial condition at large radii in our linear eigenvalue code, which is integrated backwards and  connected to the solution coming from small radii.

\section{Comparison with the results of the linear analysis}

\begin{figure}
  \begin{minipage}{0.47\textwidth}
    \centering    
    \includegraphics[width=\columnwidth]{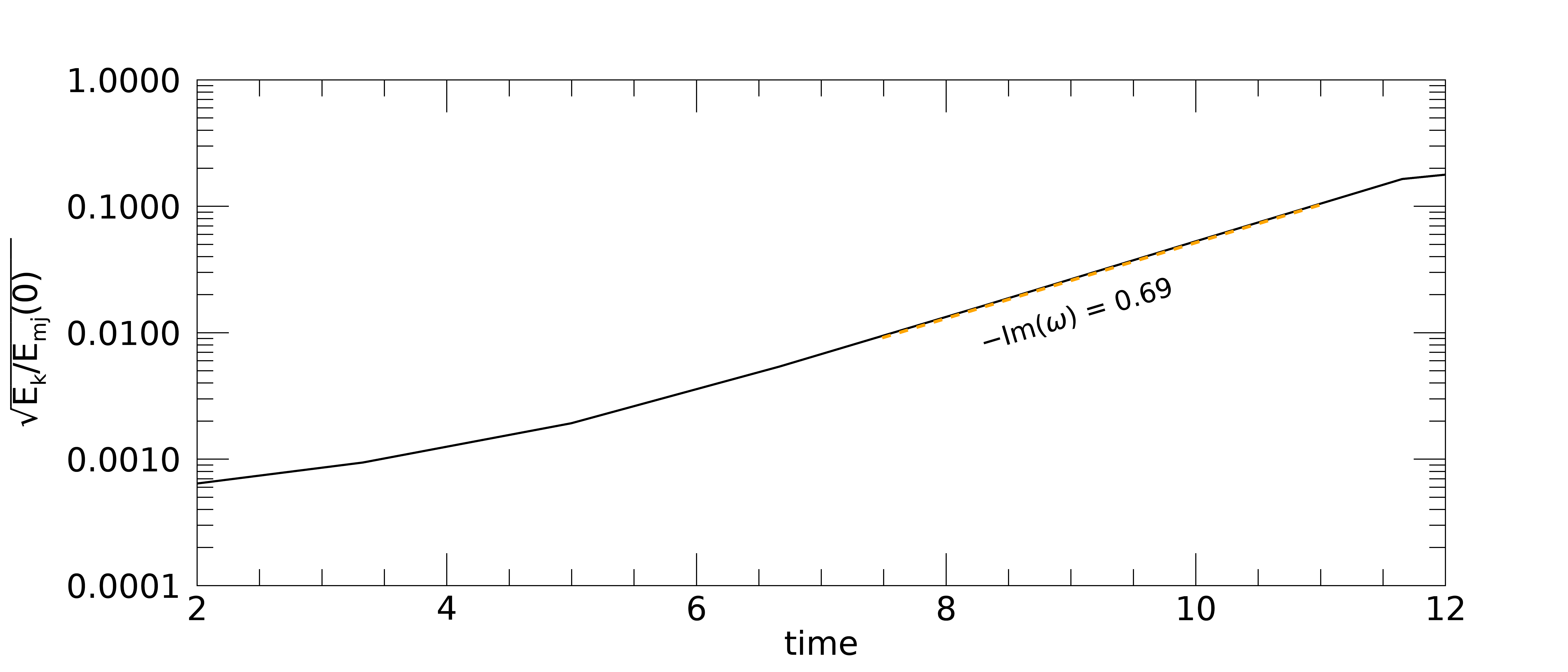}     
  \end{minipage}
  \hspace{5mm}
  \begin{minipage}{0.47\textwidth}
    \centering  
    \includegraphics[width=\columnwidth]{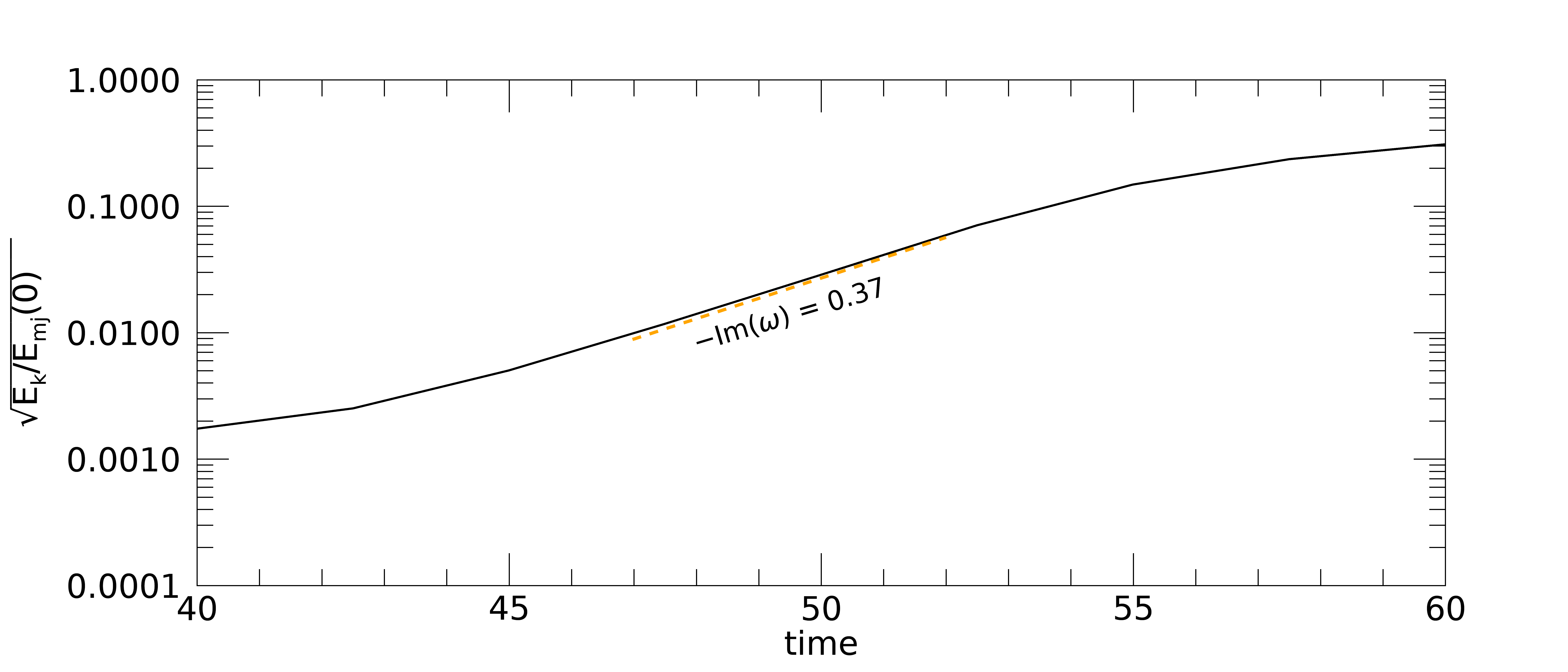}   
  \end{minipage}
  \hspace{5mm}
  \begin{minipage}{0.47\textwidth}
    \centering   
    \includegraphics[width=\columnwidth]{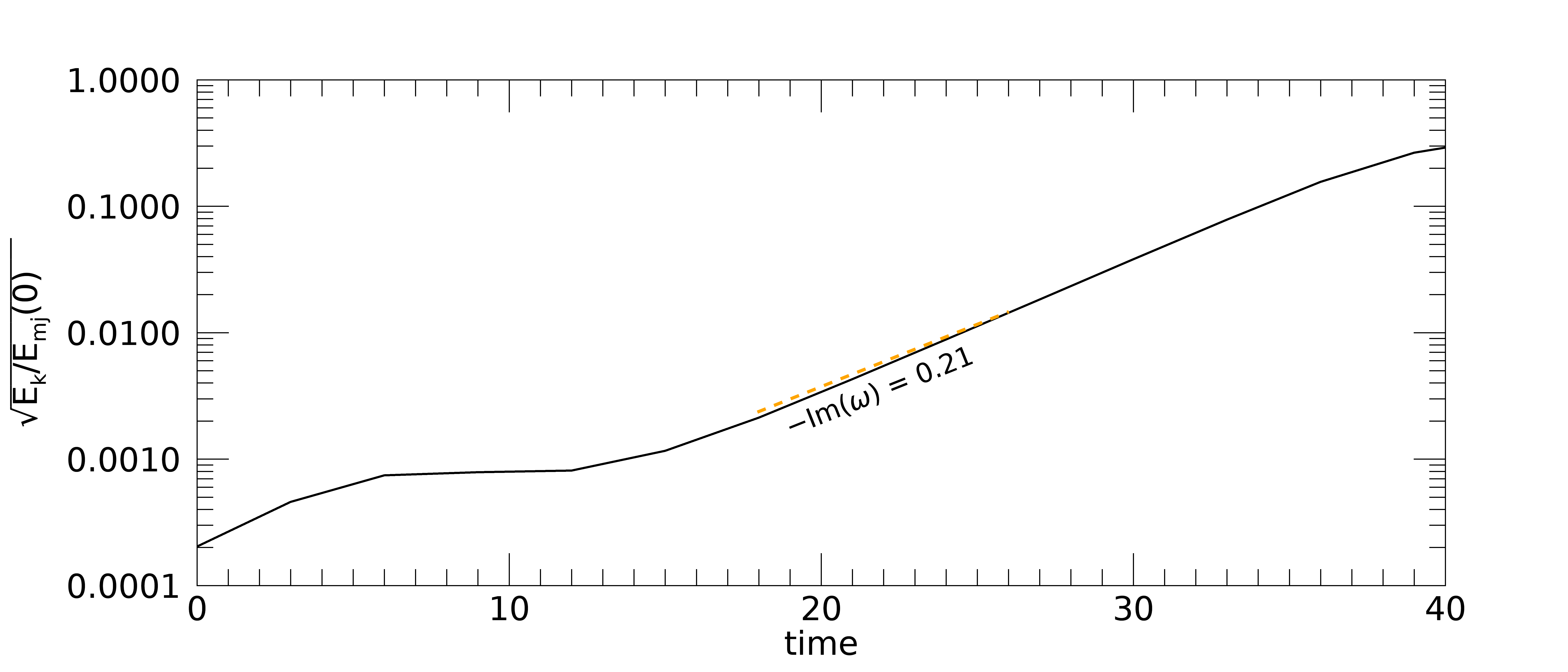}     
  \end{minipage}
    \caption{Log-linear plot of the square root of the kinetic energy normalised by the initial electromagnetic energy within the jet as a function of time for the PB (top), CHI06 (middle) and FF (bottom) simulations. The dashed lines have a slope corresponding to the expected growth rate associated with the most unstable dominant  mode retrieved from the linear analysis for the FF ($k_m=0.7$) and CHI06 ($k_m=1.1$) cases. The slope in the PB case corresponds to the linear growth rate at $k=4$. }
    \label{fig:lincomp}
\end{figure}

 Here we compare the results of the numerical simulations in the linear regime with the growth rate derived from the linear analysis. More precisely, in Fig. \ref{fig:lincomp} we display, for the three cases PB, CHI06 and FF,  a log-lin plot of the kinetic energy $E_K$ (normalized to the initial jet electromagnetic energy) as a function of time together with a straight line whose slope corresponds to the growth rate derived from the linear analysis in Sec. \ref{sec:linear}. Note that since we do not perturb the system with the exact eigenfunction at the beginning of the evolution, there is a transient phase during which the evolution is determined by the superposition of different modes. After this transient phase, the  mode with the highest growth rate eventually prevails, so that we can observe its exponential growth until it reaches the nonlinear regime. The time interval where this exponential growth takes place is therefore limited  by the initial transient stage at the beginning and by the nonlinear stage at the end, so it can be relatively short. For the FF and CHI06 cases, there are well defined large-scale modes (see Figs. \ref{fig:linear2}, \ref{fig:PB_3D} and \ref{fig:curr}), in particular, the dominant mode in the FF case has $k\approx 0.7$ and in the CHI06 case has $k\approx 1.1$ with the growth rates $-Im(\omega)= 0.21$ and $-Im(\omega)= 0.37$, respectively. More complex is the situation for the PB case, since for this case there is no well defined dominant large-scale mode, but the growth rate reaches the highest values and becomes almost flat at large wavenumbers $k$ (Fig. \ref{fig:linear2}), implying that many small-scale modes should dominate instead in the simulations, as  evident in Figs. \ref{fig:PB_3D} and \ref{fig:curr}. On the other hand, at larger $k$ numerical dissipation can hinder the growth of the instability, however, it is difficult to characterize  this dissipation and the associated critical wavenumber beyond which it becomes important. The slope that best fits the numerical results corresponds to a value of $k \approx 4$. Overall, the results of this comparison give a fairly good agreement between the numerical results and the growth rates obtained from the linear analysis.

\end{document}